\documentclass[%
reprint,
superscriptaddress,
twocolumn,
bibnotes,
amsmath,amssymb,
]{revtex4-2}
\usepackage{xcolor}

\usepackage{footmisc}
\usepackage{graphicx}
\usepackage{epstopdf}

\usepackage{dcolumn}
\usepackage{bm}
\usepackage{titlesec}
\titleformat{\section}{\raggedright\fontsize{12.5}{25}\bfseries}{\arabic{section}.}{1em}{}
\pdfoutput=1

\titleformat{\section}{\raggedright\fontsize{12.5}{25}\bfseries}{\arabic{section}.}{1em}{}
\pdfoutput=1

\DeclareUnicodeCharacter{0301}{\'a}
\DeclareUnicodeCharacter{00D6}{\"o}
\DeclareUnicodeCharacter{00C4}{\"a}
\DeclareUnicodeCharacter{00FC}{\"u}
\DeclareUnicodeCharacter{00DF}{\ss}

\usepackage{hyperref}

\usepackage{ulem}

\begin{document}
\pagenumbering{arabic}

\title{\fontsize{15}{19}\selectfont Photoinduced Electronic Band Dynamics and Defect-mediated Surface Potential Evolution in PdSe$_2$}

\author{Omar Abdul-Aziz}
\affiliation{II. Physikalisches Institut, Universit\"at zu K\"oln, Z\"ulpicher Stra\ss e 77, K\"oln D-50937, Germany}

\author{Manuel Tuniz}
\affiliation{Dipartimento di Fisica, Università di Trieste, via A. Valerio 2, 34127, Trieste, Italy}

\author{Wibke Bronsch}
\affiliation{Elettra - Sincrotrone Trieste S.C.p.A., Strada Statale 14 - km 163.5 in AREA Science Park, 34149 Basovizza, Trieste, Italy}

\author{Fulvio Parmigiani}
\affiliation{Elettra - Sincrotrone Trieste S.C.p.A., Strada Statale 14 - km 163.5 in AREA Science Park, 34149 Basovizza, Trieste, Italy}

\author{Federico Cilento}
\email{federico.cilento@elettra.eu}
\affiliation{Elettra - Sincrotrone Trieste S.C.p.A., Strada Statale 14 - km 163.5 in AREA Science Park, 34149 Basovizza, Trieste, Italy}

\author{Daniel Wolverson}
\affiliation{Department of Physics and Centre for Photonics and Photonic Materials, University of Bath, BA2 7AY Bath, UK}

\author{Charles J. Sayers}
\affiliation{Dipartimento di Fisica, Politecnico di Milano, Piazza L. da Vinci 32, 20133 Milan, Italy}

\author{Giulio Cerullo}
\affiliation{Dipartimento di Fisica, Politecnico di Milano, Piazza L. da Vinci 32, 20133 Milan, Italy}

\affiliation{Istituto di Fotonica e Nanotecnologie, Consiglio Nazionale delle Ricerche, Piazza L. da Vinci 32, 20133 Milano, Italy}

\author{Claudia Dallera}
\affiliation{Dipartimento di Fisica, Politecnico di Milano, Piazza L. da Vinci 32, 20133 Milan, Italy}

\author{Ettore Carpene}
\affiliation{Istituto di Fotonica e Nanotecnologie, Consiglio Nazionale delle Ricerche, Piazza L. da Vinci 32, 20133 Milano, Italy}

\author{Paul H. M. van Loosdrecht}
\email{pvl@ph2.uni-koeln.de}
\affiliation{II. Physikalisches Institut, Universit\"at zu K\"oln, Z\"ulpicher Stra\ss e 77, K\"oln D-50937, Germany}

\author{Hamoon Hedayat}
\email{hedayat@ph2.uni-koeln.de}
\affiliation{II. Physikalisches Institut, Universit\"at zu K\"oln, Z\"ulpicher Stra\ss e 77, K\"oln D-50937, Germany}

\begin{abstract}
\section{Abstract} We use time- and angle-resolved photoemission spectroscopy (TR-ARPES) combined with density functional theory to investigate ultrafast carrier dynamics in low-symmetry layered semiconducting PdSe$_2$. The indirect bandgap is determined to be 0.55~eV. Following photoexcitation above this gap, we resolve a valence band shift and broadening, both lasting less than a picosecond, consistent with bandgap renormalization and carrier scattering, indicative of strong many-body interactions. Subsequently, hot carriers populate the conduction band minimum and are captured by defect states. A surface photovoltage (SPV) of $\sim$ 67~meV emerges, persisting for over 50~ps, driven by defect-assisted charge separation. The formation of native vacancies, promoted by the low-symmetry lattice, likely gives rise to the mid-gap states responsible for this long-lived SPV response. Detailed analysis of TR-ARPES spectra disentangles the contributions of bandgap renormalization, carrier scattering, defect states, and SPV. These findings establish PdSe$_2$ as a prototypical layered quantum material exhibiting exotic photoresponses on ultrafast timescales.

\end{abstract}
\maketitle

\section{Introduction}
Quantum materials composed of stacked van der Waals layers have attracted tremendous attention in scientific research and advanced technology owing to their distinctive quantum effects and remarkable physical properties~\cite{manzeli20172d, kumar2025exploring,setayeshmehr2021photoconversion,schneider2018two,mak2016photonics}. Among these is the noble transition metal dichalcogenide (TMDC) PdSe$_{2}$. Distinct from the widely studied TMDCs such as MoS$_2$, WS$_2$, and WSe$_2$, PdSe$_{2}$ features a unique puckered layered structure comparable to that of black phosphorus (BP)~\cite{samy2021review,oyedele2017pdse2, xia2014rediscovering, morita1986semiconducting}, and has recently been shown to induce anisotropic, gate-tunable spin–orbit coupling in two-dimensional heterostructures~\cite{sierra2025room}. Its corrugated morphology yields a low symmetry, giving rise to exotic properties such as strong in-plane anisotropy~\cite{oyedele2017pdse2}. Moreover, PdSe$_{2}$ exhibits high chemical stability compared to BP, which suffers from degradation under ambient conditions, thereby limiting its applications~\cite{long2019palladium,pi2019recent,hassan2020performance,xia2014rediscovering, wu2019highly}. Its crystallographic direction-dependent physical properties, high electron field-effect mobility, and exceptional photodetection performance, with a high quantum efficiency across the visible to mid-infrared range, make it a promising candidate for next-generation electronics, optoelectronics, spintronics, and valleytronics~\cite{lan2019penta,zeng2019controlled,deng2018strain,gu2020two,oyedele2017pdse2,liang2019high, liu2025uncooled}. Despite recent progress in uncovering various properties of PdSe$_{2}$, including anisotropic phonon response~\cite{luo2020anisotropic,abdul2024resonance}, oxygen substitution effects~\cite{liang2020performance}, linear dichroism~\cite{yu2020direct}, strong anisotropic magnetoresistance~\cite{zhu2021observation}, and pressure-induced superconductivity~\cite{soulard2004experimental,elghazali2017pressure}, its intrinsic electronic band structure and photoexcited carrier dynamics remain relatively unexplored.\newline
The features of the electronic band structure of PdSe$_{2}$ are still under debate, with studies reporting widely varying band gap values. For the bulk, estimates range from 0 to 0.52 eV~\cite{nishiyama2022bandgap,nishiyama2022quantitative,sun2015electronic, oyedele2017pdse2,wei2022layer}, while for the monolayer, values between 1.32 and 1.43 eV have been reported~\cite{zhao2020electronic,kuklin2019quasiparticle, sun2015electronic,li2021phonon}. These discrepancies are largely attributed to the choice of theoretical models and experimental techniques.
Angle-resolved photoemission spectroscopy (ARPES) has recently been applied to investigate the band structure of bulk PdSe$_{2}$~\cite{cattelan2021site,gu2022low,chung2024dark}. Initial ARPES studies identified the valence band (VB) maximum (VBM) at the Brillouin zone (BZ) center. The absence of conduction band (CB) states at or below the Fermi level confirmed the semiconducting character~\cite{cattelan2021site}, in contrast to earlier predictions suggesting a semimetallic nature~\cite{oyedele2017pdse2}. Additional ARPES investigations revealed a highly dispersive electronic band along the interlayer direction, indicative of strong interlayer coupling~\cite{ryu2022direct}. Under in situ electron doping via alkali metal deposition, the CB minimum (CBM) became observable in ARPES measurements, and the indirect band gap was estimated to be approximately 0.36 eV~\cite{gu2022low}. However, such doping techniques may alter the intrinsic electronic structure, an effect that has also been observed in other materials, such as in BP, where band modulation and Stark effects have been reported~\cite{kim2015observation}. More recently, dark states in the VB of PdSe$_{2}$ were discovered as a result of sublattice interference, rendering certain bands invisible to ARPES across the BZs~\cite{chung2024dark}. These interference effects stem from multiple glide-mirror symmetries associated with the PdSe$_{2}$ complex sublattice structure and may be relevant to other quantum materials. Additionally, temperature- and photon energy-dependent valence band asymmetries have been observed~\cite{shan2025matrix}, adding further complexity to the interpretation of static ARPES results.\\
Although these ARPES-based studies offer valuable insights into the electronic properties of PdSe$_{2}$, a more direct method to access the unoccupied states, and the bandgap is time-resolved ARPES (TR-ARPES), which in addition enables a study of nonequilibrium carrier dynamics~\cite{boschini2024time,aeschlimann2025time,hedayat2021ultrafast}. This technique, widely employed in the study of quantum materials, enables simultaneous energy- and momentum-resolved tracking of electronic states in the time-domain following optical excitation with an ultrashort laser pulse. It also allows the investigation of transient processes such as band structure renormalization, carrier relaxation, and coherent phonon coupling, key to the understanding light–matter interactions.
In PdSe$_{2}$, investigations into these ultrafast processes have so far been limited, focusing primarily on its optical response. These studies, mainly using pump–probe THz spectroscopy, revealed a 4.3 THz coherent phonon mode coupled to interlayer charge carriers~\cite{li2021phonon}. TR-ARPES has emerged as an effective tool in revealing the detailed electronic response in analogous layered materials such as BP, where it has uncovered phenomena such as photoinduced band gap renormalization (BGR), broadening, and Stark effects~\cite{hedayat2021non,kremer2021ultrafast,chen2018band,roth2019photocarrier}. Moreover, TR-ARPES has enabled the exploration of the influence of defect states on carrier dynamics~\cite{gierster2021ultrafast,weinelt2005electronic,weinelt2004dynamics}, as well as surface photovoltage (SPV) effects and band bending dynamics under photoexcitation~\cite{ciocys2019tracking, yang2014electron,ulstrup2015ramifications, ciocys2020manipulating}.\\
Recent theoretical studies have shown that intrinsic point defects in PdSe$_2$, particularly Se vacancies and Pd+Se vacancy complexes, can introduce localized in-gap states and magnetic moments. Owing to the PdSe$_2$ characteristic five-fold morphology, Se vacancies are energetically favorable and highly mobile in both lateral and vertical directions, while Pd vacancies exhibit significantly higher diffusion barriers and remain largely immobile. This anisotropic defect behavior supports the formation of weakly dispersive, real-space localized states that enable charge trapping and internal field modulation in PdSe$_2$~\cite{kuklin2021point,jena2023evidence,nguyen20183d}.\\
In this work, we combine TR-ARPES with density functional theory (DFT) calculations to investigate the ultrafast carrier dynamics in pristine bulk PdSe$_2$. This approach allows direct population of the conduction band and a precise determination of the indirect band gap. Moreover, we resolve the energy- and momentum-dependent relaxation pathways of photoexcited carriers, enabling us to disentangle transient valence band renormalization from the emergence of a pronounced surface photovoltage (SPV) induced by optical excitation.
\section{Results}
\subsection{Electronic structure and sample characterization}
\begin{figure*}[t]
\centering
\includegraphics[width=1\textwidth]{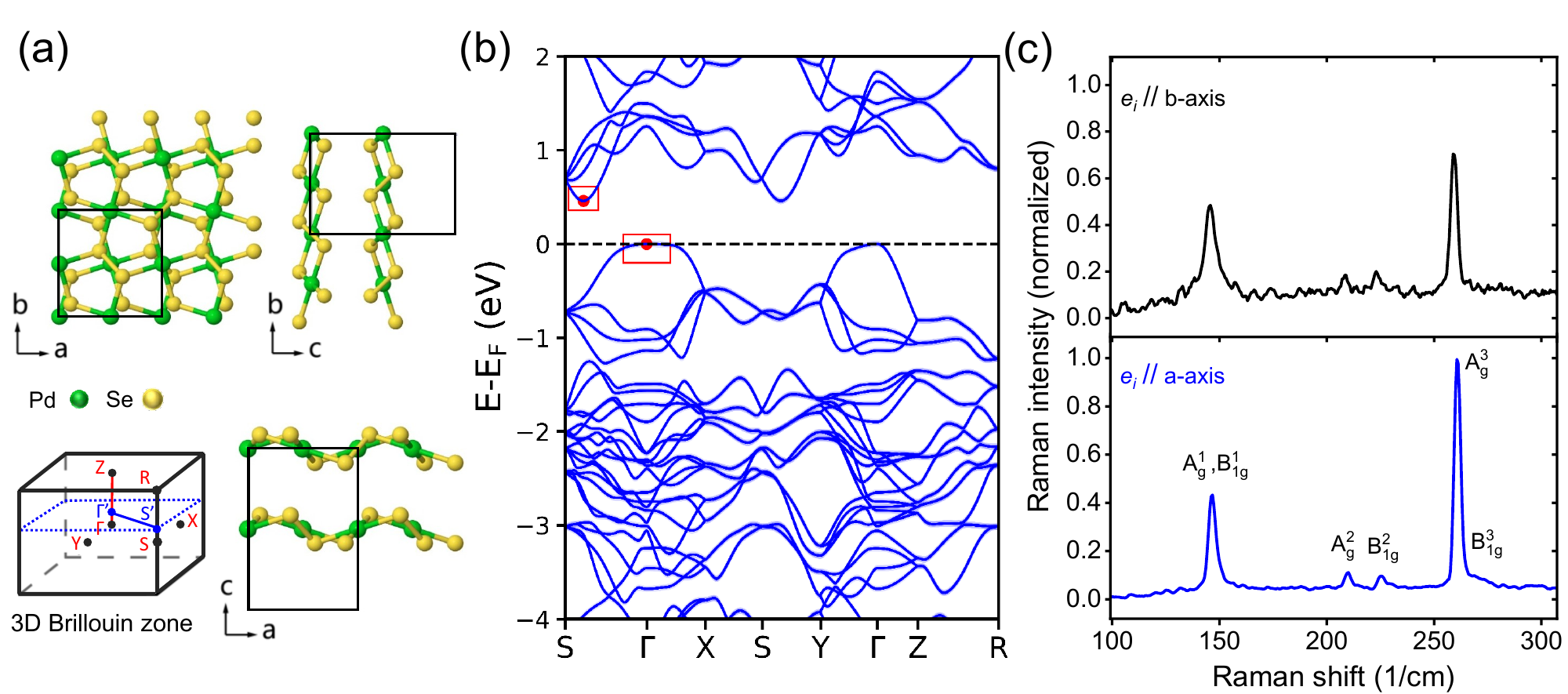}
\caption{\textbf{Electronic structure of PdSe$_2$.} (a) Top and side views of the PdSe$_2$ crystal structure (upper part), along with a 3D representation of the Brillouin zone showing high-symmetry points. The lower-right part illustrates the corrugated layer structure of PdSe$_2$. (b) $k$-resolved density of states (DOS) calculated using density functional theory (DFT) along high-symmetry paths in reciprocal space. Red squares and dots highlight the upper valence band and lower conduction band positions. (c) Raman spectrum of bulk PdSe$_2$ measured at room temperature using 532~nm excitation, with the polarization of the incident light parallel to the \textit{a}-axis (lower panel) and to the \textit{b}-axis (upper panel).
}
\label{f1}
\end{figure*}
The crystallographic structure of PdSe$_{2}$ is shown in Fig.~\ref{f1}a. In its most stable configuration, PdSe$_{2}$ adopts an orthorhombic crystal structure, classified under the Pbca space group (No. 61), with lattice parameters: \( a = 5.741 \, \text{\AA} \), \( b = 5.806 \, \text{\AA} \), and \( c = 7.691 \, \text{\AA} \), determined via X-ray diffraction~\cite{gronvold1957crystal}. The bulk unit cell consists of two PdSe$_{2}$ layers separated by a van der Waals gap, while each monolayer has a thickness of 5.2 \text{\AA}~\cite{zhang2024quantum,oyedele2017pdse2}. In the top-view projection, the monolayer forms a puckered network of non-hexagonal rings similar to the corrugated layered structure of BP~\cite{samy2021review}, as depicted in Fig.~\ref{f1}a. Structurally, it comprises three atomic subplanes, with Pd atoms centrally positioned and covalently bonded to four Se atoms in the upper and lower subplanes~\cite{oyedele2017pdse2}.\\
Based on theoretical calculations, we focus on the $\Gamma$-S direction, where the band dispersion is maximal. This direction enables us to probe the VBM and CBM (red dots in Fig.~\ref{f1}b), consistent with other first-principles DFT calculations and prior experimental results~\cite{gu2022low,cattelan2021site}.\\
The samples were initially characterized by Raman spectroscopy using 532 nm excitation. To prevent potential sample damage, the laser power was maintained below 200~\(\mu\text{W}\), with a spot size of 20~\(\mu\text{m}\). To eliminate surface oxidation, samples were cleaved via micromechanical cleavage using adhesive tape prior to measurement. Raman experiments were performed in parallel geometry with the incident and scattered lights polarized along both the a-axis or the b-axis, as shown in the lower and upper panels of Fig.~\ref{f1}c, respectively.  The Raman spectra of PdSe$_{2}$, shown in Fig.~\ref{f1}c, reveal six major peaks corresponding to in-plane and out-of-plane \(\mathrm{A_g}\) and \(\mathrm{B_{1g}}\) vibrational modes. Specifically, the \(\mathrm{A_g}\)-symmetry modes are observed at 147.7, 209.4, and 260.6 cm$^{-1}$, while the \(\mathrm{B_{1g}}\)-symmetry modes are found at 149.5, 225.9 and 271.5 cm$^{-1}$.  Based on the strong in-plane anisotropy, the sample crystal was aligned along either the a-axis or b-axis, as shown in Fig.~\ref{f1}c. The observed Raman modes and their relative intensities are consistent with those reported for bulk PdSe$_{2}$ in the literature~\cite{oyedele2017pdse2,abdul2024resonance}.

\subsection{Determination of the indirect bandgap}
\begin{figure*}
\includegraphics[width=1\textwidth]{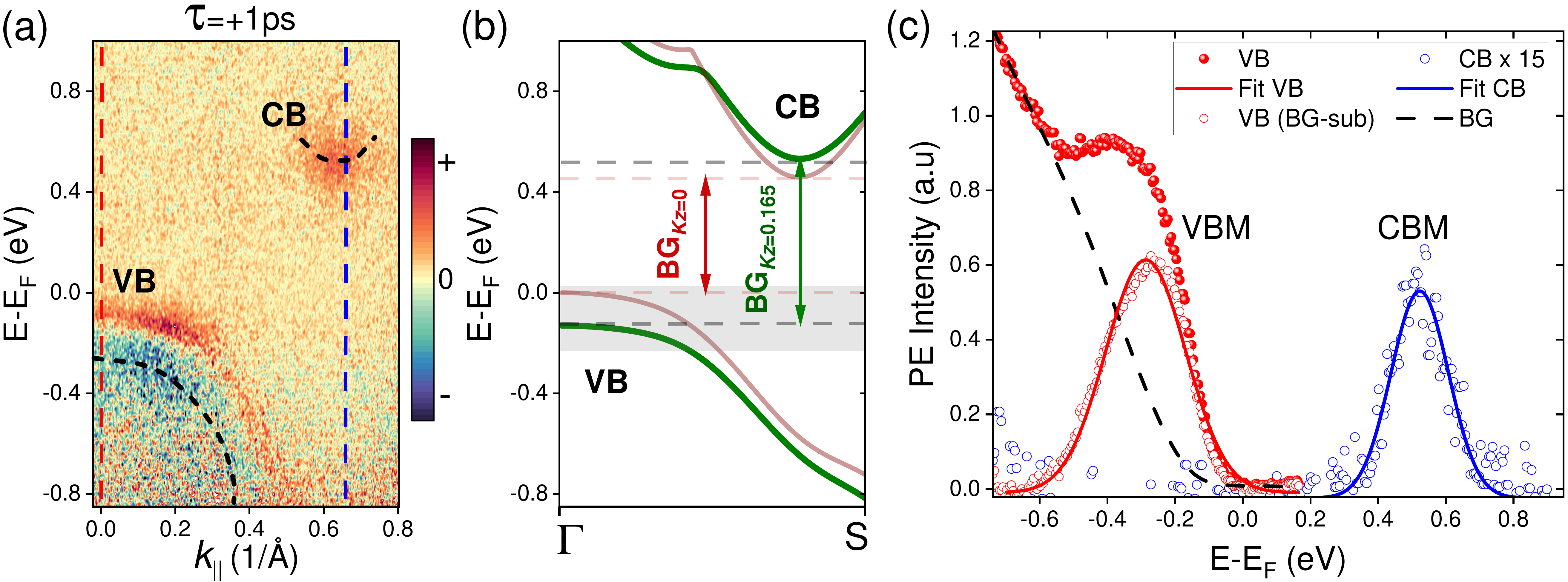}
\caption{ \textbf{Determination of the indirect bandgap.} (a) Differential ARPES map of PdSe$_{2}$ along the $\Gamma^\prime$-S$^\prime$ direction at a pump-probe delay of 1 ps, showing conduction band (CB) population. Dark brown indicates enhancement, while dark turquoise represents depletion. The CBM and VBM are observed simultaneously, and the black dashed lines serve as guides to the eye for the band dispersions. Vertical red and blue dashed lines mark the energy cuts at the VB and CB, respectively. (b) DFT calculations for the highest VB and lowest CB along the $\Gamma$-S axis at $k_z = 0$ (brown bands) and along $\Gamma^\prime$-S$^\prime$ at $k_z = 0.165$ plane (green bands) across the BZ (for details see the text). This figure demonstrates the strong $k_z$ dependence as the band gap changes from 0.45 eV up to 0.66 eV for $k_z = 0$ and $k_z = 0.165$, respectively. The grey shaded region indicates the energy range of defect-induced states obtained from DFT calculations (see Supplementary Information). (c) EDCs of the VB (red) and CB (blue), extracted along the corresponding dashed lines in panel (a) with an integration width of 0.04 Å$^{-1}$ from the absolute intensity of the TR-ARPES signal. The EDC intensity corresponding to the CB has been multiplied by a factor of 15. The bandgap at this $k_z$ point is approximately 0.80~eV (from the VB peak to the CB peak), whereas the indirect band gap of PdSe$_2$  is calculated to be 0.55~eV; for details, see the main text.}
\label{f2}
\end{figure*}

For TR-ARPES measurements on PdSe$_{2}$, the sample was oriented along the ${\Gamma-\text{S}}$ high-symmetry direction (see Fig.~\ref{f1}b) to enable simultaneous detection of the VBM and CBM within a single acquisition. Fig.~\ref{f2}a presents a TR-ARPES map of bulk PdSe$_{2}$, as acquired along the ${\Gamma-\text{S}}$ direction with 10.8 eV probe photon energy, following excitation by a pump pulse of 1.2 eV photon energy with a duration of 300 fs. This map was obtained by subtracting the photoemission spectrum recorded before pump arrival from the spectrum at a 1 ps pump-probe delay. Dark brown regions represent photo-induced increase in spectral weight, while dark turquoise indicates spectral weight depletion. \\

For comparison, Fig.~\ref{f2}b schematically shows where the CBM is estimated to be near the Brillouin zone corner.
The band dispersion, shown in Fig.~\ref{f2}a simultaneously reveals the VBM and CBM, directly confirming the indirect band gap of bulk PdSe$_{2}$. At 1 ps, the TR-ARPES spectrum exhibits clear optical excitation across the band gap, leading to electron population in the CB. The VBM is located at the $\Gamma$-point. The VB exhibits a pronounced response at 1 ps, with intensity depletion below and enhancement above the VBM, which will be discussed later. The CBM is observed 0.54 eV above the Fermi level in the corresponding constant energy map, slightly away from the Brillouin zone corners, in good agreement with our DFT results shown in Fig.~\ref{f1}b. \\  
The energy distribution curves (EDCs) of the VB and CB were extracted from the corresponding red and blue dashed lines at the $\Gamma$-point and at 0.65 Å\(^{-1}\), with an integration width of 0.04 Å\(^{-1}\), shown in Fig.~\ref{f2}c. We used a fitting method combining a Shirley-Sherwood, slope and baseline components to subtract the inelastic background. The resulting spectra enabled accurate extraction of the valence band peak and energy gap (see Supplementary Information for further details). The VBM is located at $-0.26$ eV below the Fermi level at the $\Gamma$-point. We note that the probe photon energy in our photoemission experiment does not access the $\Gamma$–S plane at $k_z = 0$, where the indirect band gap reaches its minimum, but instead probes finite out-of-plane momenta corresponding to the $\Gamma^\prime$–S$^\prime$ plane (see Fig.~\ref{f2}a). To determine the indirect band gap of PdSe$_2$, we combine TR-ARPES measurements with the relative $k_z$-dependent dispersion trends from DFT. Our TR-ARPES experiment utilized a probe photon energy of 10.80~eV. Based on the known periodicity of $\Gamma \rightarrow Z \rightarrow \Gamma$ features observed in photon-energy-dependent ARPES \cite{gu2022low}, the chosen probe photon energy of TRARPES selects an out-of-plane momentum estimated as $k_z = 0.162$ on a normalized scale where $k_z = 0$ represents the $\Gamma$ point and $k_z = 0.5$ represents the $Z$ point. We measured the band gap experimentally at this $k_z$ value as 0.80~eV (i.e., between $\Gamma^\prime{}$ and S$^\prime{}$, corresponding to the symmetry points at $k_z = 0.162$ plane). While DFT may not yield the precise absolute band gap, it accurately models the relative variation of the gap along $k_z$. Therefore, we used the DFT-calculated dispersion trend to extrapolate our experimental value from $k_z = 0.162$ back to the $\Gamma$ point ($k_z = 0$). This procedure yields an estimated indirect band gap at $\Gamma$ of 0.55~eV. This value is grounded in the experimental measurement and relies only on the relative $k_z$ dispersion predicted by DFT (see Supplementary Information for details).\\  

\subsection{Photoinduced band dynamics}

\begin{figure*}
\includegraphics[width=1\textwidth]{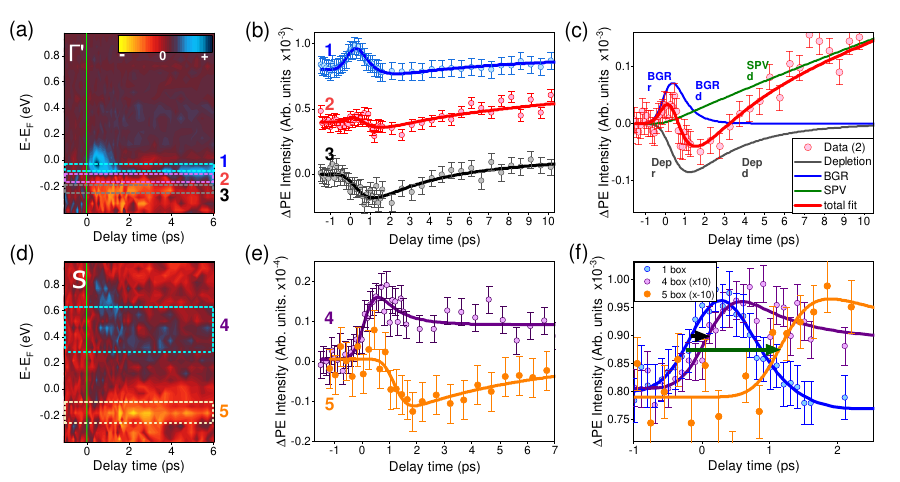}
\caption{\textbf{Photoinduced bandgap renormalization, depletion, and surface photovoltage.} (a) Differential ARPES map at the $\Gamma^\prime{}$ point shows transient VB evolution after photoexcitation. (b) Pump-probe traces or the difference in photoemission (PE) intensity at selected energy regions (1–3), each integrated over a 30 meV window, highlight distinct dynamical behaviors. The error bars scale with the fitting uncertainty. (c) Multi-exponential fit reveals components attributed to band gap renormalization, depletion (Dep), and SPV, with shared time constants. (d) Differential ARPES at $\sim$0.65~\AA$^{-1}$ shows CB electron population and below-E$\mathrm{_F}$ depletion. (e) The dynamics of the conduction band minimum (region 4) and the depletion region (region 5), integrated over energy widths of 300 and 130 meV respectively, exhibit asymmetric behavior and delayed hole trapping. (f) Comparative rise times indicate ultrafast VB broadening, CB rise $\sim$ 0.3~ps, and delayed hole depletion ($\sim$ 1~ps), represented by the black and brown arrows, respectively.}
\label{f3}
\end{figure*} 
Figure~\ref{f3}a displays the differential photoemission intensity near the $\Gamma^\prime$ point, obtained by subtracting spectra measured before pump excitation ($\tau<0$~ps) from those measured after ($\tau>0$~ps). This map reveals transient changes, including intensity increases and decreases, around the VBM. Several processes can influence the VB lineshape after excitation, such as BGR and SPV (affecting energy position), spectral depletion (reducing intensity), and increased carrier scattering (causing band broadening). To quantify these transient modifications, we analyze the photoemission intensity dynamics at different energies. We assume that these processes modify the measured lineshape of the VB EDC. Each process evolves with a characteristic timescale and influences the VB dynamics, but can affect distinct energy regions of the VB differently, leading to varying increases or decreases in the photoemission intensity across the spectrum. To disentangle these contributions and track the VB dynamics, the differential signal (Fig.~\ref{f3}a) is integrated over three representative energy regions below $E_F$ (dotted boxes, $\sim$30~meV width). This procedure yields the temporal traces shown in Fig.~\ref{f3}b. To model these temporal traces, we employ a function incorporating three contributions characterized by distinct timescales, attributed to BGR, Depletion (Dep), and SPV processes. Using this simple set of assumptions, we are able to successfully reproduce the spectral weight dynamics of the VB. The time-dependent photoemission intensity $PI_{(n)}(t)$ in the $n^\text{th}$ energy region is given by:
\begin{equation}
\begin{split}
PI_{(n)}(t) = \bigg\{\bigg[ & A_{(n)}^\mathrm{BGR} \left( 1 - e^{-t/\tau_\mathrm{r}^\mathrm{BGR}} \right) e^{-t/\tau_\mathrm{d}^\mathrm{BGR}} \\
& + A_{(n)}^\mathrm{Dep} \left( 1 - e^{-t/\tau_\mathrm{r}^\mathrm{Dep}} \right) e^{-t/\tau_\mathrm{d}^\mathrm{Dep}} \\
& + A_{(n)}^\mathrm{SPV} e^{-t/\tau_\mathrm{d}^\mathrm{SPV}} \bigg] \times \mathcal{H}(t) \bigg\} \otimes \mathcal{R}(t),
\end{split}
\label{eq:fit_dynamics}
\end{equation}
where $A_{(n)}$ denotes the amplitude of each component in the $n^\text{th}$ energy region, and $\tau_\mathrm{r}$, $\tau_\mathrm{d}$ represent the rise and decay time constants, respectively. The function $\mathcal{H}(t)$ is the Heaviside step function, and $\mathcal{R}(t)$ accounts for the Gaussian instrument response. A key assumption is that all regions share the same time constants ($\tau_\mathrm{r}$, $\tau_\mathrm{d}$) for each process, while the amplitudes ($A_{(n)}$) vary with energy. The extracted timescale values from fitting are summarized in Fig.~\ref{f3}c: BGR rise time $\tau_\mathrm{r}^{\mathrm{BGR}}$ is limited by time resolution; decay time $\tau_\mathrm{d}^{\mathrm{BGR}} = 0.68 \pm 0.35$~ps; depletion rise $\tau_\mathrm{r}^{\mathrm{Dep}} = 0.62 \pm 0.41$~ps; depletion decay $\tau_\mathrm{d}^{\mathrm{Dep}} = 2.76 \pm 0.77$~ps; and a long-lived SPV decay $\tau_\mathrm{d}^{\mathrm{SPV}} = 59.75 \pm 12.73$~ps (additional discussion is provided in the Supplementary Information.)\\
Fig.~\ref{f3}d presents the temporal evolution of the ARPES intensity at 0.65 Å\(^{-1}\), corresponding to the momentum of the CBM (blue dashed line in Fig.~\ref{f2}a). The differential map reveals an increase in spectral intensity following optical excitation, indicating the accumulation of electrons at the CBM. Interestingly, a reduction in photoemission intensity is seen below \( E_F \), yet both static ARPES measurements and DFT calculations do not predict any states at this energy and momentum. Fig.~\ref{f3}e shows the dynamics of the energy regions above and below \( E_F \), corresponding to dashed boxes 4 and 5 in Fig.~\ref{f3}d. After an initial fast decay of approximately 0.8~ps, attributed to intraband thermalization and scattering processes, the CBM exhibits negligible further dynamics and remains populated over long timescales. In contrast, the depletion observed in region 5 follows a distinct dynamical behavior compared to the VBM at the $\Gamma^\prime$-S$^\prime$ point as it exhibits a delayed rise and a significantly slower recovery, persisting for more than 6~ps. This indicates that region 5 is not associated with the intrinsic valence band dynamics but rather reflects a separate relaxation channel. This behavior is inconsistent with strong secondary electron contributions, which are generated almost instantaneously through inelastic scattering and lack both temporal structure and energy selectivity. The distinct delay in the signal onset (see the arrow in Fig.~\ref{f3}f), its localized energy character (region 5), and its picosecond dynamics collectively support an interpretation based on hole trapping into mid-gap states introduced by intrinsic defects.\\
Fig.~\ref{f3}f compares the rise times of BGR (\( \tau_\mathrm{r}^\mathrm{BGR} \)) with those of the CB and the depletion in region 5. The \(\sim\)0.3~ps delay in CB rise reflects intraband and interband scattering processes, whereas the depletion emerges only after \(\sim\)1~ps, supporting a scenario of defect-assisted carrier trapping. Our DFT calculations, together with the observed violation of the selection rules in Raman spectra, support the presence of such defect states (for details, see Supplementary Information). The role of these mid-gap states in generating a long-lived SPV is further discussed in the next section. Such defect-induced states, most likely native Se or Pd vacancy states that lie within the band gap, are generally weakly dispersive and localized in real space. They may not be detectable using static ARPES due to their low cross-section and occupation. However, following photoexcitation, hole redistribution, matrix element changes, or enhanced spectral weight due to altered band renormalization can reveal their presence. This differential analysis, comparing pumped and unpumped spectra, isolates the transient pump-induced dynamics by suppressing the static background, thereby enhancing their detectability. The timescales of trapping dynamics are in line with observations across various 2D TMDCs (e.g., MoS$_2$, WS$_2$), where this phenomenon is attributed to phonon-assisted trapping or defect capture of the Auger type~\cite{gao2021defect}.\\

\subsection{Surface photovoltage and transient band Renormalization}

\begin{figure*}
\includegraphics[width=1\textwidth]{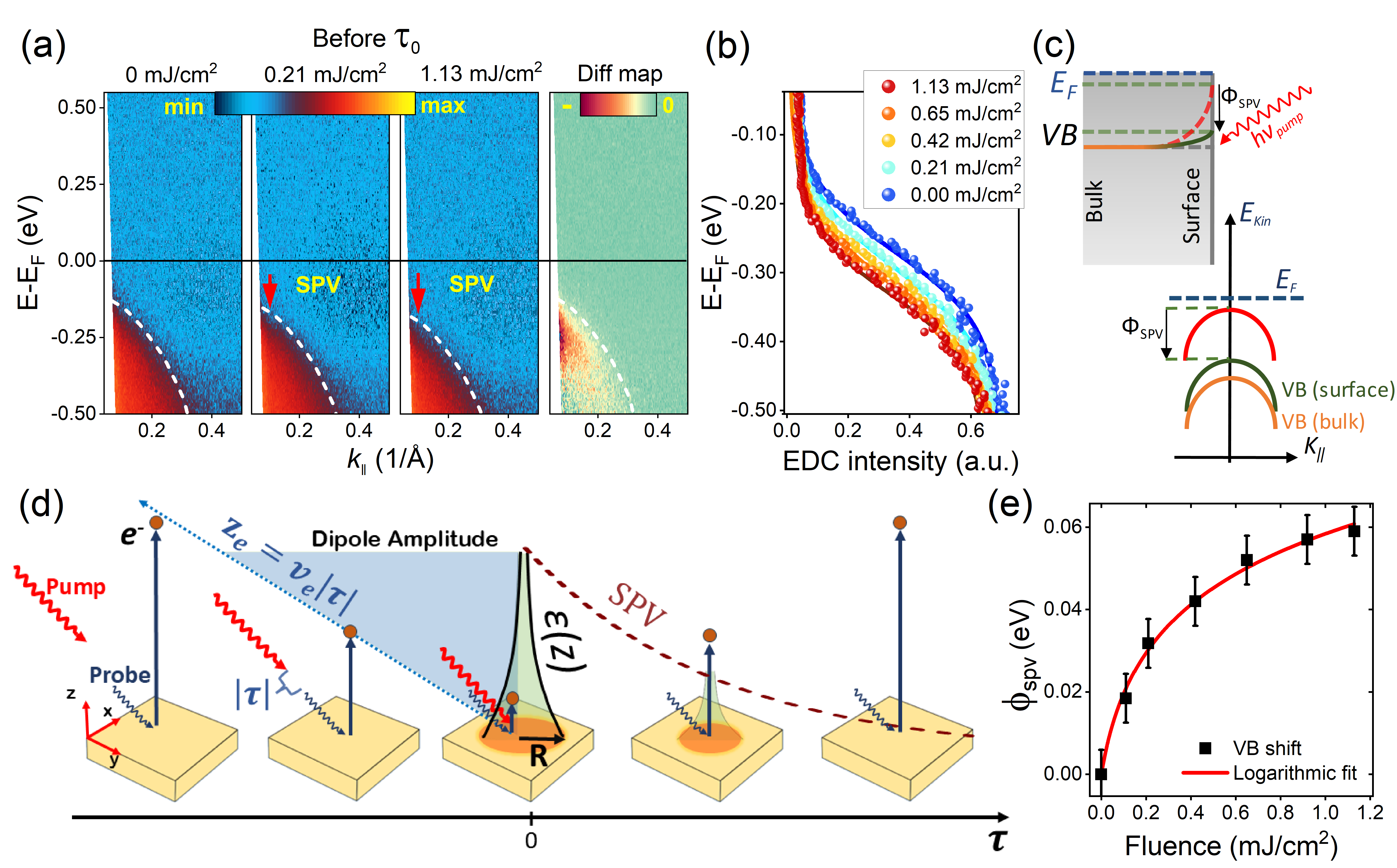}
\caption{\textbf{Analysis of the surface photovoltage effect.} (a) ARPES maps at three different pump fluences and a negative pump-probe delay $\sim -1 \, \text{ps}$, along with a differential ARPES map at high pump fluence highlighting the negative shift of the top of the VB. (b) EDCs  (bullets) and their fits (solid lines) near the top of the valence band, integrated around $\sim$ 0.1 Å$^{-1}$, for five different excitation powers, showing a clear band shift with increasing intensity. The shift of the spectra is the result of SPV. Interpretation of the positive- and negative-delay dynamics in VB. (c) The upper inset illustrates the valence and Fermi bands of the bulk and at the surface. The dashed red line represents the band bending potential, \( \Phi_{\text{SPV}} \) (surface photo-voltage). The lower inset in this panel sketches the expected ARPES spectrum. (d) A comparison of the event sequences. The black lines perpendicular to the sample surface represent the SPV-induced electric field. (e) SPV versus pump fluence. Experimental data (black marks) were fitted by the test function Eq.(2) (solid red line).
}
\label{f4}
\end{figure*}
\begin{figure}
    \centering
    \includegraphics[width=\columnwidth]{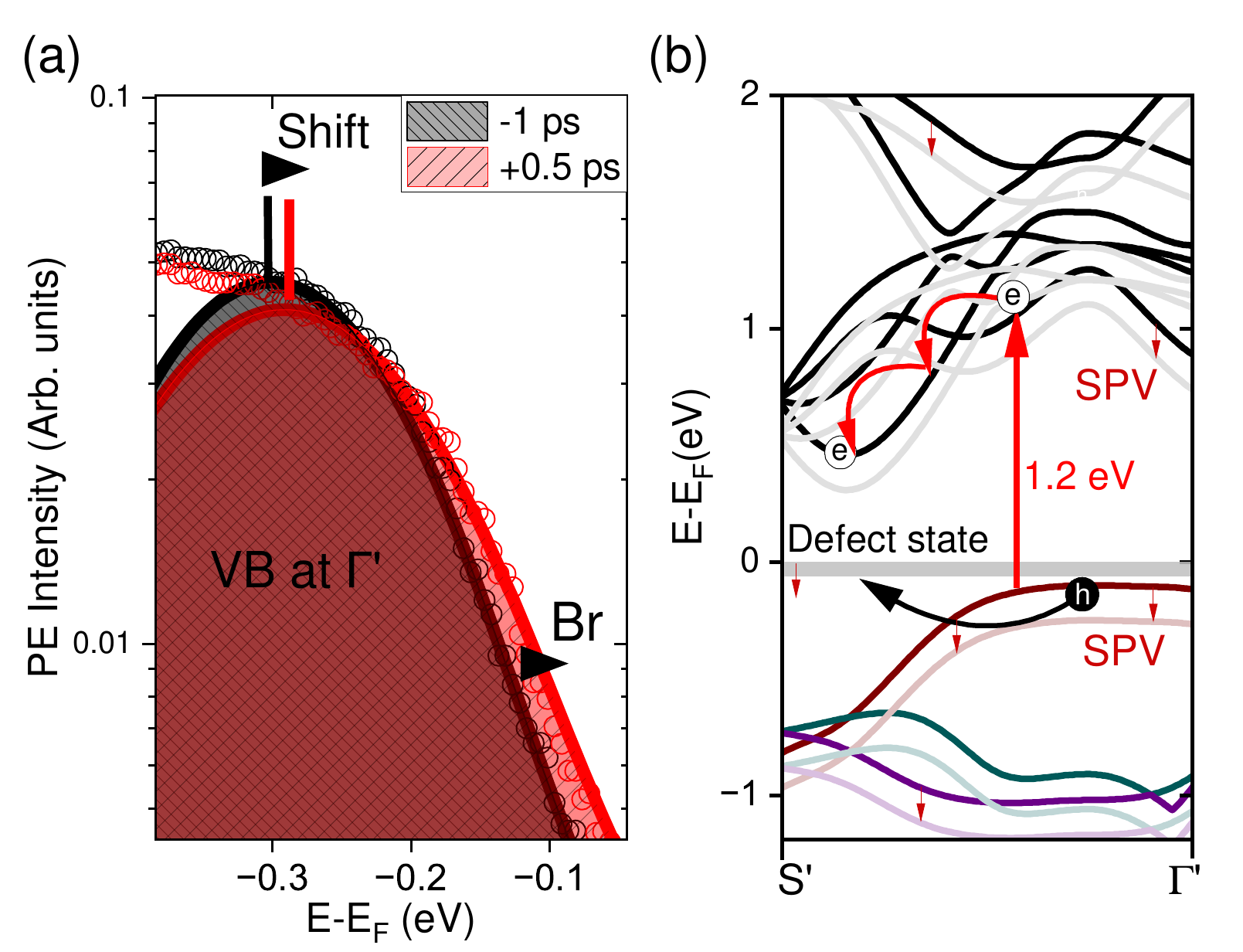}
    \caption{ \textbf{Valence band dynamics and carrier scattering pathways.}
    (a)  VB energy distribution curves at -1 ps and 0.5 ps delays (black and red dots). The solid lines show background-subtracted spectra fitted with Gaussian profiles, revealing a valence band shift and broadening indication of band gap renormalization.  (b) Schematic of carrier redistribution after photoexcitation with a 1.2 eV pump: the top VB is directly excited, and electrons reach the CBM in about ~300 fs. Holes are transferred to localized surface defect states, triggering the SPV process. 
    }
    \label{f5}
\end{figure}

Fig.~\ref{f4} reveals the emergence and evolution of SPV in PdSe$_2$, in agreement with previous studies on other materials~\cite{hedayat2021non,kremer2021ultrafast,ciocys2019tracking,yang2014electron,ulstrup2015ramifications,ciocys2020manipulating}. Fig.~\ref{f4}a shows the VB dispersion for three pump fluences measured 1 ps before the pump pulse arrival ($\mathrm{\tau_{0}}$). As the pump fluence increases, the entire VB shifts rigidly toward higher binding energies. This systematic shift is a hallmark of SPV formation: photoexcited electron–hole pairs screen the equilibrium surface band bending by partially neutralizing the space-charge field, thereby raising the local vacuum level and shifting the entire band structure to higher binding energy in the photoemission spectrum~\cite{hedayat2021non,kremer2021ultrafast}. Fig.~\ref{f4}b presents the corresponding EDCs of the VB, showing a rigid shift without significant broadening upon increasing pump fluence. This confirms that the shift is electrostatic in origin and not due to heating or lifetime effects. The fluence-dependent energy shift directly reflects the transient flattening of the band bending potential via SPV. Fig.~\ref{f4}c sketches the band alignment before and after pump excitation. In equilibrium, intrinsic defects generate upward band bending at the surface. The arrival of the pump pulse creates a population of carriers in the near-surface region, which screen the built-in field and reduce the band bending by an amount $\Phi_{\mathrm{SPV}}$; this manifests as a transient, rigid shift of both the VB and core levels toward higher binding energy~\cite{ciocys2019tracking,yang2014electron}.\\
Fig.~\ref{f4}d illustrates the event sequence from negative to positive delay configurations. For positive delays, the pump pulse initially induces the SPV and the corresponding electric field. This field influences the propagating photoelectrons, as shown in the green shaded region. The electric field decays exponentially, after which the electrons are no longer affected by the SPV. The sketch explains why a SPV effect is also observed at negative delays. The probe pulse arrives before the pump, ejecting electrons in the unperturbed band structure. For negative delays ($\tau < 0$~ps), the electron is photoemitted before the pump pulse reaches the sample. However, while the electron propagates away from the surface, the pump pulse arrives and generates a transient SPV by spatially separating photocarriers within the near-surface depletion region. This separation creates a macroscopic dipole layer whose associated electric field extends into the vacuum. As the emitted electron continues its flight, it enters this evolving field and experiences an acceleration, gaining additional kinetic energy. The energy gain $\Delta E_{\mathrm{kin}}(d)$ depends on the local dipole potential at the electron's distance $Z_e = v_e|\tau|$ from the surface, $\Delta E_{\mathrm{kin}}(d) = e\phi(d)$ where $\phi(d)$ is the potential created by the dipole field. To model this, prior works~\cite{ciocys2019tracking,yang2014electron,ciocys2020manipulating} use a disk-shaped dipole layer of radius $R$, representing the illuminated pump region. The potential at a point along the axis perpendicular to the disk (i.e., along the photoelectron path) is given by, $ \phi(d) = \phi_0 \left(1 - \frac{d}{\sqrt{d^2 + R^2}} \right)$. At long negative delays, this effect disappears, as we measure no difference between the unpumped spectrum and the spectrum at delay $-200$~ps, consistent with the absence of a SPV. Fig.~\ref{f4}e shows the extracted SPV potential $\Phi_{\mathrm{SPV}}$ as a function of pump fluence. The maximum SPV shift is approximately 67~meV, yielding a surface defect density on the order of $10^{12}$–$10^{13}$~cm$^{-2}$ (see Supplementary Information for estimation details). The data follow a phenomenological logarithmic saturation trend, described by:
\begin{equation}
\Phi_{\mathrm{SPV}} = \alpha \ln \left(1 + \frac{\rho}{\rho_0} \right),
\end{equation}
where $\alpha = 20.8$ meV and $\rho_0 = 63.38 \times 10^{-3} \,\mathrm{cm}^2/\mathrm{m}\mathrm{J}$ are the fitting parameters. This trend indicates that SPV generation saturates at higher fluences, consistent with carrier screening and depletion-layer filling effects in semiconductors~\cite{ulstrup2015ramifications,hedayat2021non}. Therefore, our results confirm that PdSe$_2$ exhibits a robust SPV response under ultrafast photoexcitation. This effect is attributed to intrinsic defect states near the surface facilitating long-lived charge separation. Our DFT calculations (see Supplementary Information) align with theoretical studies indicate that selenium vacancies in PdSe$_2$ are mobile and can migrate towards the surface, forming localized in-gap states~\cite{jena2023evidence}. Given their propensity to be negatively charged, these surface vacancies are expected to create an initial upward band bending. Therefore, the most plausible mechanism consistent with our observations is that upon photoexcitation, the trapping of photogenerated holes by these states partially neutralizes the surface charge. This leads to a flattening of the band bending, which manifests as the observed SPV.\\
We finally focus on the analysis of the transient spectral lineshapes to derive the potential VB energy shift and broadening after photoexcitation. In Fig.~\ref{f5}a, the solid lines depict Gaussian fits to the background-subtracted EDCs at delays of $-1$~ps and $0.5$~ps, allowing for accurate determination of variations in linewidth and central energy (see SI for more information). Therefore, one can determine the photo-driven VB dynamics, showing modest VB broadening of 22~meV and an energy shift of approximately 15~meV, which we attribute to BGR and increased carrier scattering.

\section{Discussion}
Fig.~\ref{f5}(b) schematically illustrates the underlying mechanism: upon 1.2~eV excitation, electrons are promoted from the VB to the CBs, where they thermalize to the CBM within $\sim$ 300 fs. The photogenerated holes are trapped in localized surface defect states, resulting in spatial charge separation, flattening of the band bending and the formation of a transient electric field responsible for the SPV. This separation generates a macroscopic dipole layer, which modifies the kinetic energy of photoelectrons depending on the pump-probe delay. These findings confirm that photoexcitation of carriers results in ultrafast BGR and broadening. Further studies may help clarify whether the observed effects are related to the Stark effect reported in BP~\cite{chen2018band}.

Our study unveils the ultrafast photoinduced charge dynamics in semiconducting PdSe$_{2}$, revealing a rich interplay between transient band dynamics and SPV formation. Time-resolved ARPES measurements directly capture the population of conduction band states and renormalization of the valence band within sub-picosecond timescales. These effects are accompanied by a pronounced, long-lived surface photovoltage, driven by defect-assisted hole trapping and macroscopic dipole layer formation. The indirect band gap is determined to be 0.55 eV.  From a detailed investigation of the TR-ARPES results, we disentangle the contributions of BGR and band broadening. The SPV shift in PdSe$_{2}$ exceeds 67~meV. The combination of air-stable low-symmetry structure, strong many-body effects, and defect-mediated surface fields positions PdSe$_{2}$ as a compelling platform for light-tunable carrier separation and ultrafast surface potential engineering. These features may open pathways for developing high-speed optoelectronic switches and photogated transistors leveraging surface charge dynamics. Crucially, the demonstrated influence of native vacancies in shaping mid-gap state dynamics and enabling long-lived SPV underscores the pivotal role of intrinsic defect landscapes in governing the nonequilibrium electronic responses of quantum materials.

\section{Methods}

\subsection{Time-resolved and angle-resolved photoemission spectroscopy}
Bulk PdSe$_{2}$ samples were obtained from HQ Graphene~\cite{hqgraphene}. Prior to TR-ARPES measurements, the samples were cleaved in-situ at room temperature in ultrahigh vacuum (UHV) conditions, with a pressure better than \( \sim2\times10^{-10} \, \text{mbar} \). The orientation of the sample was determined by low-energy electron diffraction (LEED).
TR-ARPES measurements were conducted at the T-ReX laboratory operated by the Elettra Sincrotrone Trieste, S.C.p.A.,  using the setup described in Ref.~\cite{peli2020time}, based on a 200 kHz Coherent Monaco laser system. The output centered at 1.2 eV was used as the pump. The ninth harmonics (10.8 eV) of the laser was used as the pobe and was p-polarized. The pump fluence was tunable, and a variable delay ($\tau$) was applied to the sample relative to the probe pulse. The pump and probe beams were focused to spot sizes of approximately 300 \(\mu\text{m}\) and 200 \(\mu\text{m}\) respectively, and impinged on the sample surface at a 30-degree angle. The experimental time, energy, and angular resolutions were better than 800 fs, 50 meV, and 0.1 degrees, respectively.\\
The sample temperature was maintained at $\sim$100 K during the measurements. The photoemitted electrons were detected by a hemispherical analyzer, with an acceptance angle of \( \pm 15^\circ \). 
\subsection{Density Functional Theory}
Density functional theory (DFT) calculations of the band structure were carried out using the plane wave DFT code Quantum Espresso~\cite{giannozzi2009quantum,giannozzi2017advanced}. In the DFT calculations presented in Fig.~\ref{f1}b, ultrasoft pseudopotentials with the PBE exchange-correlation functional~\cite{perdew1996generalized} were used with a kinetic energy cutoff of 680 eV and a Monkhorst-Pack~\cite{monkhorst1976special} $k$-point density of $8 \times 8 \times 6$ for the primitive unit cell. The structures were relaxed to achieve forces of order $10^{-3}$ eV\AA$^{-1}$ and pressures less than $\pm 0.2$ kbar. Results were compared to (i) calculations using the regularized SCAN meta-GGA exchange-correlation functional~\cite{bartok2019regularized} along with the rVV10 kernel for the inclusion of van der Waals forces between layers~\cite{sabatini2013nonlocal} and to (ii) Heyd-Scuseria-Ernzerhof hybrid functional~\cite{heyd2003hybrid} calculations with the use of Wannier90~\cite{mostofi2014updated} for interpolation. The trends of the band edge positions at the $\Gamma^\prime$ and S$^\prime$ points as a function of $k_z$ were found to be similar across all levels of theory.\\

\section*{Data Availability}

The TR-ARPES data that support the findings of this study are openly available in \url{https://doi.org/10.5281/zenodo.XXXXXXX}. The DFT calculations were performed using the open-source Quantum Espresso package. No custom code was developed for this work. 

\section*{Acknowledgements\label{ack}}
The authors acknowledge financial support by the Deutsche Forschungsgemeinschaft (DFG) through project German Research Foundation via project No. 277146847 - CRC 1238: Control and Dynamics of quantum Materials. Computational work was performed on the University of Bath’s High Performance Computing Facility and was supported by the EU Horizon 2020 OCRE/GEANT project “Cloud funding for research”. G.C. acknowledges financial support by the European Union’s NextGenerationEU Programme with the I-PHOQS Infrastructure [IR0000016, ID D2B8D520, CUP B53C22001750006] "Integrated infrastructure initiative in Photonic and Quantum Sciences". C.J.S. and G.C. acknowledge support from the Horizon Europe EIC Pathfinder Open program under grant agreement no. 101130384 (QUONDENSATE).

\section*{Author Contributions}
 O.A.A., M.T., W.B., F.C., C.J.S., E.C., and H.H. performed the experimental measurements. O.A.A. analyzed the data under the supervision of H.H. and P.v.L. D.W. performed the DFT calculations. All authors, O.A.A., M.T., W.B., F.P., F.C., D.W. C.J.S., G.C., C.D., E.C., P.v.L., and H.H. contributed to the discussion and interpretation of the results. The manuscript was written by O.A.A. and H.H. with input from all authors.

\subsection*{COMPETING INTERESTS}
The authors declare no competing financial or non-financial interests.

\subsection*{ADDITIONAL INFORMATION}
\textbf{Supplementary Information} The online version contains supplementary material
available at https://doi.org/XXX.

\def\bibsection{\section*{~\refname}} 
\bibliography{main}

\providecommand{\noopsort}[1]{}\providecommand{\singleletter}[1]{#1}
\begin{thebibliography}{70}%
\makeatletter
\providecommand \@ifxundefined [1]{%
 \@ifx{#1\undefined}
}%
\providecommand \@ifnum [1]{%
 \ifnum #1\expandafter \@firstoftwo
 \else \expandafter \@secondoftwo
 \fi
}%
\providecommand \@ifx [1]{%
 \ifx #1\expandafter \@firstoftwo
 \else \expandafter \@secondoftwo
 \fi
}%
\providecommand \natexlab [1]{#1}%
\providecommand \enquote  [1]{``#1''}%
\providecommand \bibnamefont  [1]{#1}%
\providecommand \bibfnamefont [1]{#1}%
\providecommand \citenamefont [1]{#1}%
\providecommand \href@noop [0]{\@secondoftwo}%
\providecommand \href [0]{\begingroup \@sanitize@url \@href}%
\providecommand \@href[1]{\@@startlink{#1}\@@href}%
\providecommand \@@href[1]{\endgroup#1\@@endlink}%
\providecommand \@sanitize@url [0]{\catcode `\\12\catcode `\$12\catcode `\&12\catcode `\#12\catcode `\^12\catcode `\_12\catcode `\%12\relax}%
\providecommand \@@startlink[1]{}%
\providecommand \@@endlink[0]{}%
\providecommand \url  [0]{\begingroup\@sanitize@url \@url }%
\providecommand \@url [1]{\endgroup\@href {#1}{\urlprefix }}%
\providecommand \urlprefix  [0]{URL }%
\providecommand \Eprint [0]{\href }%
\providecommand \doibase [0]{https://doi.org/}%
\providecommand \selectlanguage [0]{\@gobble}%
\providecommand \bibinfo  [0]{\@secondoftwo}%
\providecommand \bibfield  [0]{\@secondoftwo}%
\providecommand \translation [1]{[#1]}%
\providecommand \BibitemOpen [0]{}%
\providecommand \bibitemStop [0]{}%
\providecommand \bibitemNoStop [0]{.\EOS\space}%
\providecommand \EOS [0]{\spacefactor3000\relax}%
\providecommand \BibitemShut  [1]{\csname bibitem#1\endcsname}%
\let\auto@bib@innerbib\@empty
\bibitem [{\citenamefont {Manzeli}\ \emph {et~al.}(2017)\citenamefont {Manzeli}, \citenamefont {Ovchinnikov}, \citenamefont {Pasquier}, \citenamefont {Yazyev},\ and\ \citenamefont {Kis}}]{manzeli20172d}%
  \BibitemOpen
  \bibfield  {author} {\bibinfo {author} {\bibfnamefont {S.}~\bibnamefont {Manzeli}}, \bibinfo {author} {\bibfnamefont {D.}~\bibnamefont {Ovchinnikov}}, \bibinfo {author} {\bibfnamefont {D.}~\bibnamefont {Pasquier}}, \bibinfo {author} {\bibfnamefont {O.~V.}\ \bibnamefont {Yazyev}},\ and\ \bibinfo {author} {\bibfnamefont {A.}~\bibnamefont {Kis}},\ }\bibfield  {title} {\bibinfo {title} {2\text{D} transition metal dichalcogenides},\ }\href@noop {} {\bibfield  {journal} {\bibinfo  {journal} {Nature Reviews Materials}\ }\textbf {\bibinfo {volume} {2}},\ \bibinfo {pages} {1} (\bibinfo {year} {2017})}\BibitemShut {NoStop}%
\bibitem [{\citenamefont {Kumar}\ \emph {et~al.}(2025)\citenamefont {Kumar}, \citenamefont {Raut}, \citenamefont {Panda},\ and\ \citenamefont {Rashed}}]{kumar2025exploring}%
  \BibitemOpen
  \bibfield  {author} {\bibinfo {author} {\bibfnamefont {V.~P.}\ \bibnamefont {Kumar}}, \bibinfo {author} {\bibfnamefont {P.}~\bibnamefont {Raut}}, \bibinfo {author} {\bibfnamefont {D.~K.}\ \bibnamefont {Panda}},\ and\ \bibinfo {author} {\bibfnamefont {A.~N.~Z.}\ \bibnamefont {Rashed}},\ }\bibfield  {title} {\bibinfo {title} {Exploring next-generation tmdc materials: A comprehensive review of their classifications, properties, and applications},\ }\href@noop {} {\bibfield  {journal} {\bibinfo  {journal} {Silicon}\ ,\ \bibinfo {pages} {1}} (\bibinfo {year} {2025})}\BibitemShut {NoStop}%
\bibitem [{\citenamefont {Setayeshmehr}\ \emph {et~al.}(2021)\citenamefont {Setayeshmehr}, \citenamefont {Hashemi},\ and\ \citenamefont {Ansari}}]{setayeshmehr2021photoconversion}%
  \BibitemOpen
  \bibfield  {author} {\bibinfo {author} {\bibfnamefont {K.}~\bibnamefont {Setayeshmehr}}, \bibinfo {author} {\bibfnamefont {M.}~\bibnamefont {Hashemi}},\ and\ \bibinfo {author} {\bibfnamefont {N.}~\bibnamefont {Ansari}},\ }\bibfield  {title} {\bibinfo {title} {Photoconversion efficiency in atomically thin tmdc-based heterostructures},\ }\href@noop {} {\bibfield  {journal} {\bibinfo  {journal} {Optics Express}\ }\textbf {\bibinfo {volume} {29}},\ \bibinfo {pages} {32910} (\bibinfo {year} {2021})}\BibitemShut {NoStop}%
\bibitem [{\citenamefont {Schneider}\ \emph {et~al.}(2018)\citenamefont {Schneider}, \citenamefont {Glazov}, \citenamefont {Korn}, \citenamefont {H{\"o}fling},\ and\ \citenamefont {Urbaszek}}]{schneider2018two}%
  \BibitemOpen
  \bibfield  {author} {\bibinfo {author} {\bibfnamefont {C.}~\bibnamefont {Schneider}}, \bibinfo {author} {\bibfnamefont {M.~M.}\ \bibnamefont {Glazov}}, \bibinfo {author} {\bibfnamefont {T.}~\bibnamefont {Korn}}, \bibinfo {author} {\bibfnamefont {S.}~\bibnamefont {H{\"o}fling}},\ and\ \bibinfo {author} {\bibfnamefont {B.}~\bibnamefont {Urbaszek}},\ }\bibfield  {title} {\bibinfo {title} {Two-dimensional semiconductors in the regime of strong light-matter coupling},\ }\href@noop {} {\bibfield  {journal} {\bibinfo  {journal} {Nature communications}\ }\textbf {\bibinfo {volume} {9}},\ \bibinfo {pages} {2695} (\bibinfo {year} {2018})}\BibitemShut {NoStop}%
\bibitem [{\citenamefont {Mak}\ and\ \citenamefont {Shan}(2016)}]{mak2016photonics}%
  \BibitemOpen
  \bibfield  {author} {\bibinfo {author} {\bibfnamefont {K.~F.}\ \bibnamefont {Mak}}\ and\ \bibinfo {author} {\bibfnamefont {J.}~\bibnamefont {Shan}},\ }\bibfield  {title} {\bibinfo {title} {Photonics and optoelectronics of 2\text{D} semiconductor transition metal dichalcogenides},\ }\href@noop {} {\bibfield  {journal} {\bibinfo  {journal} {Nature Photonics}\ }\textbf {\bibinfo {volume} {10}},\ \bibinfo {pages} {216} (\bibinfo {year} {2016})}\BibitemShut {NoStop}%
\bibitem [{\citenamefont {Samy}\ \emph {et~al.}(2021)\citenamefont {Samy}, \citenamefont {Zeng}, \citenamefont {Birowosuto},\ and\ \citenamefont {El~Moutaouakil}}]{samy2021review}%
  \BibitemOpen
  \bibfield  {author} {\bibinfo {author} {\bibfnamefont {O.}~\bibnamefont {Samy}}, \bibinfo {author} {\bibfnamefont {S.}~\bibnamefont {Zeng}}, \bibinfo {author} {\bibfnamefont {M.~D.}\ \bibnamefont {Birowosuto}},\ and\ \bibinfo {author} {\bibfnamefont {A.}~\bibnamefont {El~Moutaouakil}},\ }\bibfield  {title} {\bibinfo {title} {A review on {MoS}$_{2}$ properties, synthesis, sensing applications and challenges},\ }\href@noop {} {\bibfield  {journal} {\bibinfo  {journal} {Crystals}\ }\textbf {\bibinfo {volume} {11}},\ \bibinfo {pages} {355} (\bibinfo {year} {2021})}\BibitemShut {NoStop}%
\bibitem [{\citenamefont {Oyedele}\ \emph {et~al.}(2017)\citenamefont {Oyedele}, \citenamefont {Yang}, \citenamefont {Liang}, \citenamefont {Puretzky}, \citenamefont {Wang}, \citenamefont {Zhang}, \citenamefont {Yu}, \citenamefont {Pudasaini}, \citenamefont {Ghosh}, \citenamefont {Liu} \emph {et~al.}}]{oyedele2017pdse2}%
  \BibitemOpen
  \bibfield  {author} {\bibinfo {author} {\bibfnamefont {A.~D.}\ \bibnamefont {Oyedele}}, \bibinfo {author} {\bibfnamefont {S.}~\bibnamefont {Yang}}, \bibinfo {author} {\bibfnamefont {L.}~\bibnamefont {Liang}}, \bibinfo {author} {\bibfnamefont {A.~A.}\ \bibnamefont {Puretzky}}, \bibinfo {author} {\bibfnamefont {K.}~\bibnamefont {Wang}}, \bibinfo {author} {\bibfnamefont {J.}~\bibnamefont {Zhang}}, \bibinfo {author} {\bibfnamefont {P.}~\bibnamefont {Yu}}, \bibinfo {author} {\bibfnamefont {P.~R.}\ \bibnamefont {Pudasaini}}, \bibinfo {author} {\bibfnamefont {A.~W.}\ \bibnamefont {Ghosh}}, \bibinfo {author} {\bibfnamefont {Z.}~\bibnamefont {Liu}}, \emph {et~al.},\ }\bibfield  {title} {\bibinfo {title} {{PdSe}$_{2}$: pentagonal two-dimensional layers with high air stability for electronics},\ }\href@noop {} {\bibfield  {journal} {\bibinfo  {journal} {Journal of the American Chemical Society}\ }\textbf {\bibinfo {volume} {139}},\ \bibinfo {pages} {14090} (\bibinfo {year} {2017})}\BibitemShut {NoStop}%
\bibitem [{\citenamefont {Xia}\ \emph {et~al.}(2014)\citenamefont {Xia}, \citenamefont {Wang},\ and\ \citenamefont {Jia}}]{xia2014rediscovering}%
  \BibitemOpen
  \bibfield  {author} {\bibinfo {author} {\bibfnamefont {F.}~\bibnamefont {Xia}}, \bibinfo {author} {\bibfnamefont {H.}~\bibnamefont {Wang}},\ and\ \bibinfo {author} {\bibfnamefont {Y.}~\bibnamefont {Jia}},\ }\bibfield  {title} {\bibinfo {title} {Rediscovering black phosphorus as an anisotropic layered material for optoelectronics and electronics},\ }\href@noop {} {\bibfield  {journal} {\bibinfo  {journal} {Nature communications}\ }\textbf {\bibinfo {volume} {5}},\ \bibinfo {pages} {4458} (\bibinfo {year} {2014})}\BibitemShut {NoStop}%
\bibitem [{\citenamefont {Morita}(1986)}]{morita1986semiconducting}%
  \BibitemOpen
  \bibfield  {author} {\bibinfo {author} {\bibfnamefont {A.}~\bibnamefont {Morita}},\ }\bibfield  {title} {\bibinfo {title} {Semiconducting black phosphorus},\ }\href@noop {} {\bibfield  {journal} {\bibinfo  {journal} {Applied Physics A}\ }\textbf {\bibinfo {volume} {39}},\ \bibinfo {pages} {227} (\bibinfo {year} {1986})}\BibitemShut {NoStop}%
\bibitem [{\citenamefont {Sierra}\ \emph {et~al.}(2025)\citenamefont {Sierra}, \citenamefont {Sv{\v{e}}tl{\'\i}k}, \citenamefont {Savero~Torres}, \citenamefont {Camosi}, \citenamefont {Herling}, \citenamefont {Guillet}, \citenamefont {Xu}, \citenamefont {Reparaz}, \citenamefont {Marinova}, \citenamefont {Dimitrov} \emph {et~al.}}]{sierra2025room}%
  \BibitemOpen
  \bibfield  {author} {\bibinfo {author} {\bibfnamefont {J.~F.}\ \bibnamefont {Sierra}}, \bibinfo {author} {\bibfnamefont {J.}~\bibnamefont {Sv{\v{e}}tl{\'\i}k}}, \bibinfo {author} {\bibfnamefont {W.}~\bibnamefont {Savero~Torres}}, \bibinfo {author} {\bibfnamefont {L.}~\bibnamefont {Camosi}}, \bibinfo {author} {\bibfnamefont {F.}~\bibnamefont {Herling}}, \bibinfo {author} {\bibfnamefont {T.}~\bibnamefont {Guillet}}, \bibinfo {author} {\bibfnamefont {K.}~\bibnamefont {Xu}}, \bibinfo {author} {\bibfnamefont {J.~S.}\ \bibnamefont {Reparaz}}, \bibinfo {author} {\bibfnamefont {V.}~\bibnamefont {Marinova}}, \bibinfo {author} {\bibfnamefont {D.}~\bibnamefont {Dimitrov}}, \emph {et~al.},\ }\bibfield  {title} {\bibinfo {title} {Room-temperature anisotropic in-plane spin dynamics in graphene induced by {PdSe}$_{2}$ proximity},\ }\href@noop {} {\bibfield  {journal} {\bibinfo  {journal} {Nature Materials}\ ,\ \bibinfo {pages} {1}} (\bibinfo {year} {2025})}\BibitemShut {NoStop}%
\bibitem [{\citenamefont {Long}\ \emph {et~al.}(2019)\citenamefont {Long}, \citenamefont {Wang}, \citenamefont {Wang}, \citenamefont {Zhou}, \citenamefont {Xia}, \citenamefont {Luo}, \citenamefont {Huang}, \citenamefont {Zhang}, \citenamefont {Yan}, \citenamefont {Fan} \emph {et~al.}}]{long2019palladium}%
  \BibitemOpen
  \bibfield  {author} {\bibinfo {author} {\bibfnamefont {M.}~\bibnamefont {Long}}, \bibinfo {author} {\bibfnamefont {Y.}~\bibnamefont {Wang}}, \bibinfo {author} {\bibfnamefont {P.}~\bibnamefont {Wang}}, \bibinfo {author} {\bibfnamefont {X.}~\bibnamefont {Zhou}}, \bibinfo {author} {\bibfnamefont {H.}~\bibnamefont {Xia}}, \bibinfo {author} {\bibfnamefont {C.}~\bibnamefont {Luo}}, \bibinfo {author} {\bibfnamefont {S.}~\bibnamefont {Huang}}, \bibinfo {author} {\bibfnamefont {G.}~\bibnamefont {Zhang}}, \bibinfo {author} {\bibfnamefont {H.}~\bibnamefont {Yan}}, \bibinfo {author} {\bibfnamefont {Z.}~\bibnamefont {Fan}}, \emph {et~al.},\ }\bibfield  {title} {\bibinfo {title} {Palladium diselenide long-wavelength infrared photodetector with high sensitivity and stability},\ }\href@noop {} {\bibfield  {journal} {\bibinfo  {journal} {Acs Nano}\ }\textbf {\bibinfo {volume} {13}},\ \bibinfo {pages} {2511} (\bibinfo {year} {2019})}\BibitemShut {NoStop}%
\bibitem [{\citenamefont {Pi}\ \emph {et~al.}(2019)\citenamefont {Pi}, \citenamefont {Li}, \citenamefont {Liu}, \citenamefont {Zhang}, \citenamefont {Li},\ and\ \citenamefont {Zhai}}]{pi2019recent}%
  \BibitemOpen
  \bibfield  {author} {\bibinfo {author} {\bibfnamefont {L.}~\bibnamefont {Pi}}, \bibinfo {author} {\bibfnamefont {L.}~\bibnamefont {Li}}, \bibinfo {author} {\bibfnamefont {K.}~\bibnamefont {Liu}}, \bibinfo {author} {\bibfnamefont {Q.}~\bibnamefont {Zhang}}, \bibinfo {author} {\bibfnamefont {H.}~\bibnamefont {Li}},\ and\ \bibinfo {author} {\bibfnamefont {T.}~\bibnamefont {Zhai}},\ }\bibfield  {title} {\bibinfo {title} {Recent progress on 2\text{D} noble-transition-metal dichalcogenides},\ }\href@noop {} {\bibfield  {journal} {\bibinfo  {journal} {Advanced Functional Materials}\ }\textbf {\bibinfo {volume} {29}},\ \bibinfo {pages} {1904932} (\bibinfo {year} {2019})}\BibitemShut {NoStop}%
\bibitem [{\citenamefont {Hassan}\ \emph {et~al.}(2020)\citenamefont {Hassan}, \citenamefont {Guo},\ and\ \citenamefont {Wang}}]{hassan2020performance}%
  \BibitemOpen
  \bibfield  {author} {\bibinfo {author} {\bibfnamefont {A.}~\bibnamefont {Hassan}}, \bibinfo {author} {\bibfnamefont {Y.}~\bibnamefont {Guo}},\ and\ \bibinfo {author} {\bibfnamefont {Q.}~\bibnamefont {Wang}},\ }\bibfield  {title} {\bibinfo {title} {Performance of the pentagonal {PdSe}$_{2}$ sheet as a channel material in contact with metal surfaces and graphene},\ }\href@noop {} {\bibfield  {journal} {\bibinfo  {journal} {ACS Applied Electronic Materials}\ }\textbf {\bibinfo {volume} {2}},\ \bibinfo {pages} {2535} (\bibinfo {year} {2020})}\BibitemShut {NoStop}%
\bibitem [{\citenamefont {Wu}\ \emph {et~al.}(2019)\citenamefont {Wu}, \citenamefont {Guo}, \citenamefont {Du}, \citenamefont {Xia}, \citenamefont {Zeng}, \citenamefont {Tian}, \citenamefont {Shi}, \citenamefont {Tian}, \citenamefont {Li}, \citenamefont {Tsang} \emph {et~al.}}]{wu2019highly}%
  \BibitemOpen
  \bibfield  {author} {\bibinfo {author} {\bibfnamefont {D.}~\bibnamefont {Wu}}, \bibinfo {author} {\bibfnamefont {J.}~\bibnamefont {Guo}}, \bibinfo {author} {\bibfnamefont {J.}~\bibnamefont {Du}}, \bibinfo {author} {\bibfnamefont {C.}~\bibnamefont {Xia}}, \bibinfo {author} {\bibfnamefont {L.}~\bibnamefont {Zeng}}, \bibinfo {author} {\bibfnamefont {Y.}~\bibnamefont {Tian}}, \bibinfo {author} {\bibfnamefont {Z.}~\bibnamefont {Shi}}, \bibinfo {author} {\bibfnamefont {Y.}~\bibnamefont {Tian}}, \bibinfo {author} {\bibfnamefont {X.~J.}\ \bibnamefont {Li}}, \bibinfo {author} {\bibfnamefont {Y.~H.}\ \bibnamefont {Tsang}}, \emph {et~al.},\ }\bibfield  {title} {\bibinfo {title} {Highly polarization-sensitive, broadband, self-powered photodetector based on graphene {PdSe}$_{2}$ germanium heterojunction},\ }\href@noop {} {\bibfield  {journal} {\bibinfo  {journal} {ACS nano}\ }\textbf {\bibinfo {volume} {13}},\ \bibinfo {pages} {9907} (\bibinfo {year} {2019})}\BibitemShut {NoStop}%
\bibitem [{\citenamefont {Lan}\ \emph {et~al.}(2019)\citenamefont {Lan}, \citenamefont {Chen}, \citenamefont {Hu}, \citenamefont {Cheng},\ and\ \citenamefont {Chen}}]{lan2019penta}%
  \BibitemOpen
  \bibfield  {author} {\bibinfo {author} {\bibfnamefont {Y.-S.}\ \bibnamefont {Lan}}, \bibinfo {author} {\bibfnamefont {X.-R.}\ \bibnamefont {Chen}}, \bibinfo {author} {\bibfnamefont {C.-E.}\ \bibnamefont {Hu}}, \bibinfo {author} {\bibfnamefont {Y.}~\bibnamefont {Cheng}},\ and\ \bibinfo {author} {\bibfnamefont {Q.-F.}\ \bibnamefont {Chen}},\ }\bibfield  {title} {\bibinfo {title} {Penta-{PdX}$_{2}$ {(X= S, Se, Te)} monolayers: promising anisotropic thermoelectric materials},\ }\href@noop {} {\bibfield  {journal} {\bibinfo  {journal} {Journal of Materials Chemistry A}\ }\textbf {\bibinfo {volume} {7}},\ \bibinfo {pages} {11134} (\bibinfo {year} {2019})}\BibitemShut {NoStop}%
\bibitem [{\citenamefont {Zeng}\ \emph {et~al.}(2019)\citenamefont {Zeng}, \citenamefont {Wu}, \citenamefont {Lin}, \citenamefont {Xie}, \citenamefont {Yuan}, \citenamefont {Lu}, \citenamefont {Lau}, \citenamefont {Chai}, \citenamefont {Luo}, \citenamefont {Li} \emph {et~al.}}]{zeng2019controlled}%
  \BibitemOpen
  \bibfield  {author} {\bibinfo {author} {\bibfnamefont {L.-H.}\ \bibnamefont {Zeng}}, \bibinfo {author} {\bibfnamefont {D.}~\bibnamefont {Wu}}, \bibinfo {author} {\bibfnamefont {S.-H.}\ \bibnamefont {Lin}}, \bibinfo {author} {\bibfnamefont {C.}~\bibnamefont {Xie}}, \bibinfo {author} {\bibfnamefont {H.-Y.}\ \bibnamefont {Yuan}}, \bibinfo {author} {\bibfnamefont {W.}~\bibnamefont {Lu}}, \bibinfo {author} {\bibfnamefont {S.~P.}\ \bibnamefont {Lau}}, \bibinfo {author} {\bibfnamefont {Y.}~\bibnamefont {Chai}}, \bibinfo {author} {\bibfnamefont {L.-B.}\ \bibnamefont {Luo}}, \bibinfo {author} {\bibfnamefont {Z.-J.}\ \bibnamefont {Li}}, \emph {et~al.},\ }\bibfield  {title} {\bibinfo {title} {Controlled synthesis of 2\text{D} palladium diselenide for sensitive photodetector applications},\ }\href@noop {} {\bibfield  {journal} {\bibinfo  {journal} {Advanced Functional Materials}\ }\textbf {\bibinfo {volume} {29}},\ \bibinfo {pages} {1806878} (\bibinfo {year} {2019})}\BibitemShut {NoStop}%
\bibitem [{\citenamefont {Deng}\ \emph {et~al.}(2018)\citenamefont {Deng}, \citenamefont {Li},\ and\ \citenamefont {Zhang}}]{deng2018strain}%
  \BibitemOpen
  \bibfield  {author} {\bibinfo {author} {\bibfnamefont {S.}~\bibnamefont {Deng}}, \bibinfo {author} {\bibfnamefont {L.}~\bibnamefont {Li}},\ and\ \bibinfo {author} {\bibfnamefont {Y.}~\bibnamefont {Zhang}},\ }\bibfield  {title} {\bibinfo {title} {Strain modulated electronic, mechanical, and optical properties of the monolayer {PdS}$_{2}$, {PdSe}$_{2}$, and {PtSe}$_{2}$ for tunable devices},\ }\href@noop {} {\bibfield  {journal} {\bibinfo  {journal} {ACS Applied Nano Materials}\ }\textbf {\bibinfo {volume} {1}},\ \bibinfo {pages} {1932} (\bibinfo {year} {2018})}\BibitemShut {NoStop}%
\bibitem [{\citenamefont {Gu}\ \emph {et~al.}(2020)\citenamefont {Gu}, \citenamefont {Cai}, \citenamefont {Dong}, \citenamefont {Yu}, \citenamefont {Hoffman}, \citenamefont {Liu}, \citenamefont {Oyedele}, \citenamefont {Lin}, \citenamefont {Ge}, \citenamefont {Puretzky} \emph {et~al.}}]{gu2020two}%
  \BibitemOpen
  \bibfield  {author} {\bibinfo {author} {\bibfnamefont {Y.}~\bibnamefont {Gu}}, \bibinfo {author} {\bibfnamefont {H.}~\bibnamefont {Cai}}, \bibinfo {author} {\bibfnamefont {J.}~\bibnamefont {Dong}}, \bibinfo {author} {\bibfnamefont {Y.}~\bibnamefont {Yu}}, \bibinfo {author} {\bibfnamefont {A.~N.}\ \bibnamefont {Hoffman}}, \bibinfo {author} {\bibfnamefont {C.}~\bibnamefont {Liu}}, \bibinfo {author} {\bibfnamefont {A.~D.}\ \bibnamefont {Oyedele}}, \bibinfo {author} {\bibfnamefont {Y.-C.}\ \bibnamefont {Lin}}, \bibinfo {author} {\bibfnamefont {Z.}~\bibnamefont {Ge}}, \bibinfo {author} {\bibfnamefont {A.~A.}\ \bibnamefont {Puretzky}}, \emph {et~al.},\ }\bibfield  {title} {\bibinfo {title} {Two-dimensional palladium diselenide with strong in-plane optical anisotropy and high mobility grown by chemical vapor deposition},\ }\href@noop {} {\bibfield  {journal} {\bibinfo  {journal} {Advanced Materials}\ }\textbf {\bibinfo {volume} {32}},\ \bibinfo {pages} {1906238} (\bibinfo {year} {2020})}\BibitemShut {NoStop}%
\bibitem [{\citenamefont {Liang}\ \emph {et~al.}(2019)\citenamefont {Liang}, \citenamefont {Wang}, \citenamefont {Zhang}, \citenamefont {Wei}, \citenamefont {Lim}, \citenamefont {Zhu}, \citenamefont {Hu}, \citenamefont {Wei}, \citenamefont {Lee}, \citenamefont {Sow} \emph {et~al.}}]{liang2019high}%
  \BibitemOpen
  \bibfield  {author} {\bibinfo {author} {\bibfnamefont {Q.}~\bibnamefont {Liang}}, \bibinfo {author} {\bibfnamefont {Q.}~\bibnamefont {Wang}}, \bibinfo {author} {\bibfnamefont {Q.}~\bibnamefont {Zhang}}, \bibinfo {author} {\bibfnamefont {J.}~\bibnamefont {Wei}}, \bibinfo {author} {\bibfnamefont {S.~X.}\ \bibnamefont {Lim}}, \bibinfo {author} {\bibfnamefont {R.}~\bibnamefont {Zhu}}, \bibinfo {author} {\bibfnamefont {J.}~\bibnamefont {Hu}}, \bibinfo {author} {\bibfnamefont {W.}~\bibnamefont {Wei}}, \bibinfo {author} {\bibfnamefont {C.}~\bibnamefont {Lee}}, \bibinfo {author} {\bibfnamefont {C.}~\bibnamefont {Sow}}, \emph {et~al.},\ }\bibfield  {title} {\bibinfo {title} {High-performance, room temperature, ultra-broadband photodetectors based on air-stable {PdSe}$_{2}$},\ }\href@noop {} {\bibfield  {journal} {\bibinfo  {journal} {Advanced Materials}\ }\textbf {\bibinfo {volume} {31}},\ \bibinfo {pages} {1807609} (\bibinfo {year} {2019})}\BibitemShut {NoStop}%
\bibitem [{\citenamefont {Liu}\ \emph {et~al.}(2025)\citenamefont {Liu}, \citenamefont {Qi}, \citenamefont {Zou}, \citenamefont {Zhang}, \citenamefont {Zhang}, \citenamefont {Xiang}, \citenamefont {Liu}, \citenamefont {Qin}, \citenamefont {Sun}, \citenamefont {Zheng} \emph {et~al.}}]{liu2025uncooled}%
  \BibitemOpen
  \bibfield  {author} {\bibinfo {author} {\bibfnamefont {M.}~\bibnamefont {Liu}}, \bibinfo {author} {\bibfnamefont {L.}~\bibnamefont {Qi}}, \bibinfo {author} {\bibfnamefont {Y.}~\bibnamefont {Zou}}, \bibinfo {author} {\bibfnamefont {N.}~\bibnamefont {Zhang}}, \bibinfo {author} {\bibfnamefont {F.}~\bibnamefont {Zhang}}, \bibinfo {author} {\bibfnamefont {H.}~\bibnamefont {Xiang}}, \bibinfo {author} {\bibfnamefont {Z.}~\bibnamefont {Liu}}, \bibinfo {author} {\bibfnamefont {M.}~\bibnamefont {Qin}}, \bibinfo {author} {\bibfnamefont {X.}~\bibnamefont {Sun}}, \bibinfo {author} {\bibfnamefont {Y.}~\bibnamefont {Zheng}}, \emph {et~al.},\ }\bibfield  {title} {\bibinfo {title} {Uncooled near-to long-wave-infrared polarization-sensitive photodetectors based on {MoSe}$_{2}$/{PdSe}$_{2}$ van der waals heterostructures},\ }\href@noop {} {\bibfield  {journal} {\bibinfo  {journal} {Nature Communications}\ }\textbf {\bibinfo {volume} {16}},\ \bibinfo {pages} {2774} (\bibinfo {year} {2025})}\BibitemShut {NoStop}%
\bibitem [{\citenamefont {Luo}\ \emph {et~al.}(2020)\citenamefont {Luo}, \citenamefont {Oyedele}, \citenamefont {Gu}, \citenamefont {Li}, \citenamefont {Wang}, \citenamefont {Haglund}, \citenamefont {Mandrus}, \citenamefont {Puretzky}, \citenamefont {Xiao}, \citenamefont {Liang} \emph {et~al.}}]{luo2020anisotropic}%
  \BibitemOpen
  \bibfield  {author} {\bibinfo {author} {\bibfnamefont {W.}~\bibnamefont {Luo}}, \bibinfo {author} {\bibfnamefont {A.~D.}\ \bibnamefont {Oyedele}}, \bibinfo {author} {\bibfnamefont {Y.}~\bibnamefont {Gu}}, \bibinfo {author} {\bibfnamefont {T.}~\bibnamefont {Li}}, \bibinfo {author} {\bibfnamefont {X.}~\bibnamefont {Wang}}, \bibinfo {author} {\bibfnamefont {A.~V.}\ \bibnamefont {Haglund}}, \bibinfo {author} {\bibfnamefont {D.}~\bibnamefont {Mandrus}}, \bibinfo {author} {\bibfnamefont {A.~A.}\ \bibnamefont {Puretzky}}, \bibinfo {author} {\bibfnamefont {K.}~\bibnamefont {Xiao}}, \bibinfo {author} {\bibfnamefont {L.}~\bibnamefont {Liang}}, \emph {et~al.},\ }\bibfield  {title} {\bibinfo {title} {Anisotropic phonon response of few-layer {PdSe}$_{2}$ under uniaxial strain},\ }\href@noop {} {\bibfield  {journal} {\bibinfo  {journal} {Advanced Functional Materials}\ }\textbf {\bibinfo {volume} {30}},\ \bibinfo {pages} {2003215} (\bibinfo {year} {2020})}\BibitemShut {NoStop}%
\bibitem [{\citenamefont {Abdul-Aziz}\ \emph {et~al.}(2024)\citenamefont {Abdul-Aziz}, \citenamefont {Wolverson}, \citenamefont {Sayers}, \citenamefont {Carpene}, \citenamefont {Parmigiani}, \citenamefont {Hedayat},\ and\ \citenamefont {van Loosdrecht}}]{abdul2024resonance}%
  \BibitemOpen
  \bibfield  {author} {\bibinfo {author} {\bibfnamefont {O.}~\bibnamefont {Abdul-Aziz}}, \bibinfo {author} {\bibfnamefont {D.}~\bibnamefont {Wolverson}}, \bibinfo {author} {\bibfnamefont {C.~J.}\ \bibnamefont {Sayers}}, \bibinfo {author} {\bibfnamefont {E.}~\bibnamefont {Carpene}}, \bibinfo {author} {\bibfnamefont {F.}~\bibnamefont {Parmigiani}}, \bibinfo {author} {\bibfnamefont {H.}~\bibnamefont {Hedayat}},\ and\ \bibinfo {author} {\bibfnamefont {P.~H.}\ \bibnamefont {van Loosdrecht}},\ }\bibfield  {title} {\bibinfo {title} {Resonance-induced anomalies in temperature-dependent raman scattering of {PdSe}$_{2}$},\ }\href@noop {} {\bibfield  {journal} {\bibinfo  {journal} {Journal of Materials Chemistry C}\ }\textbf {\bibinfo {volume} {12}},\ \bibinfo {pages} {11402} (\bibinfo {year} {2024})}\BibitemShut {NoStop}%
\bibitem [{\citenamefont {Liang}\ \emph {et~al.}(2020)\citenamefont {Liang}, \citenamefont {Zhang}, \citenamefont {Gou}, \citenamefont {Song}, \citenamefont {Arramel}, \citenamefont {Chen}, \citenamefont {Yang}, \citenamefont {Lim}, \citenamefont {Wang}, \citenamefont {Zhu} \emph {et~al.}}]{liang2020performance}%
  \BibitemOpen
  \bibfield  {author} {\bibinfo {author} {\bibfnamefont {Q.}~\bibnamefont {Liang}}, \bibinfo {author} {\bibfnamefont {Q.}~\bibnamefont {Zhang}}, \bibinfo {author} {\bibfnamefont {J.}~\bibnamefont {Gou}}, \bibinfo {author} {\bibfnamefont {T.}~\bibnamefont {Song}}, \bibinfo {author} {\bibnamefont {Arramel}}, \bibinfo {author} {\bibfnamefont {H.}~\bibnamefont {Chen}}, \bibinfo {author} {\bibfnamefont {M.}~\bibnamefont {Yang}}, \bibinfo {author} {\bibfnamefont {S.~X.}\ \bibnamefont {Lim}}, \bibinfo {author} {\bibfnamefont {Q.}~\bibnamefont {Wang}}, \bibinfo {author} {\bibfnamefont {R.}~\bibnamefont {Zhu}}, \emph {et~al.},\ }\bibfield  {title} {\bibinfo {title} {Performance improvement by ozone treatment of 2\text{D} {PdSe}$_{2}$},\ }\href@noop {} {\bibfield  {journal} {\bibinfo  {journal} {ACS nano}\ }\textbf {\bibinfo {volume} {14}},\ \bibinfo {pages} {5668} (\bibinfo {year} {2020})}\BibitemShut {NoStop}%
\bibitem [{\citenamefont {Yu}\ \emph {et~al.}(2020)\citenamefont {Yu}, \citenamefont {Kuang}, \citenamefont {Gao}, \citenamefont {Wang}, \citenamefont {Chen}, \citenamefont {Ding}, \citenamefont {Liu}, \citenamefont {Cong}, \citenamefont {He}, \citenamefont {Liu} \emph {et~al.}}]{yu2020direct}%
  \BibitemOpen
  \bibfield  {author} {\bibinfo {author} {\bibfnamefont {J.}~\bibnamefont {Yu}}, \bibinfo {author} {\bibfnamefont {X.}~\bibnamefont {Kuang}}, \bibinfo {author} {\bibfnamefont {Y.}~\bibnamefont {Gao}}, \bibinfo {author} {\bibfnamefont {Y.}~\bibnamefont {Wang}}, \bibinfo {author} {\bibfnamefont {K.}~\bibnamefont {Chen}}, \bibinfo {author} {\bibfnamefont {Z.}~\bibnamefont {Ding}}, \bibinfo {author} {\bibfnamefont {J.}~\bibnamefont {Liu}}, \bibinfo {author} {\bibfnamefont {C.}~\bibnamefont {Cong}}, \bibinfo {author} {\bibfnamefont {J.}~\bibnamefont {He}}, \bibinfo {author} {\bibfnamefont {Z.}~\bibnamefont {Liu}}, \emph {et~al.},\ }\bibfield  {title} {\bibinfo {title} {Direct observation of the linear dichroism transition in two-dimensional palladium diselenide},\ }\href@noop {} {\bibfield  {journal} {\bibinfo  {journal} {Nano letters}\ }\textbf {\bibinfo {volume} {20}},\ \bibinfo {pages} {1172} (\bibinfo {year} {2020})}\BibitemShut {NoStop}%
\bibitem [{\citenamefont {Zhu}\ \emph {et~al.}(2021)\citenamefont {Zhu}, \citenamefont {Gao}, \citenamefont {Liang}, \citenamefont {Hu}, \citenamefont {Wang}, \citenamefont {Qiu},\ and\ \citenamefont {Wee}}]{zhu2021observation}%
  \BibitemOpen
  \bibfield  {author} {\bibinfo {author} {\bibfnamefont {R.}~\bibnamefont {Zhu}}, \bibinfo {author} {\bibfnamefont {Z.}~\bibnamefont {Gao}}, \bibinfo {author} {\bibfnamefont {Q.}~\bibnamefont {Liang}}, \bibinfo {author} {\bibfnamefont {J.}~\bibnamefont {Hu}}, \bibinfo {author} {\bibfnamefont {J.-S.}\ \bibnamefont {Wang}}, \bibinfo {author} {\bibfnamefont {C.-W.}\ \bibnamefont {Qiu}},\ and\ \bibinfo {author} {\bibfnamefont {A.~T.~S.}\ \bibnamefont {Wee}},\ }\bibfield  {title} {\bibinfo {title} {Observation of anisotropic magnetoresistance in layered nonmagnetic semiconducting {PdSe}$_{2}$},\ }\href@noop {} {\bibfield  {journal} {\bibinfo  {journal} {ACS Applied Materials \& Interfaces}\ }\textbf {\bibinfo {volume} {13}},\ \bibinfo {pages} {37527} (\bibinfo {year} {2021})}\BibitemShut {NoStop}%
\bibitem [{\citenamefont {Soulard}\ \emph {et~al.}(2004)\citenamefont {Soulard}, \citenamefont {Rocquefelte}, \citenamefont {Petit}, \citenamefont {Evain}, \citenamefont {Jobic}, \citenamefont {Iti{\'e}}, \citenamefont {Munsch}, \citenamefont {Koo},\ and\ \citenamefont {Whangbo}}]{soulard2004experimental}%
  \BibitemOpen
  \bibfield  {author} {\bibinfo {author} {\bibfnamefont {C.}~\bibnamefont {Soulard}}, \bibinfo {author} {\bibfnamefont {X.}~\bibnamefont {Rocquefelte}}, \bibinfo {author} {\bibfnamefont {P.-E.}\ \bibnamefont {Petit}}, \bibinfo {author} {\bibfnamefont {M.}~\bibnamefont {Evain}}, \bibinfo {author} {\bibfnamefont {S.}~\bibnamefont {Jobic}}, \bibinfo {author} {\bibfnamefont {J.-P.}\ \bibnamefont {Iti{\'e}}}, \bibinfo {author} {\bibfnamefont {P.}~\bibnamefont {Munsch}}, \bibinfo {author} {\bibfnamefont {H.-J.}\ \bibnamefont {Koo}},\ and\ \bibinfo {author} {\bibfnamefont {M.-H.}\ \bibnamefont {Whangbo}},\ }\bibfield  {title} {\bibinfo {title} {Experimental and theoretical investigation on the relative stability of the {PdS}$_{2}$-and pyrite-type structures of {PdSe}$_{2}$},\ }\href@noop {} {\bibfield  {journal} {\bibinfo  {journal} {Inorganic chemistry}\ }\textbf {\bibinfo {volume} {43}},\ \bibinfo {pages} {1943} (\bibinfo {year} {2004})}\BibitemShut {NoStop}%
\bibitem [{\citenamefont {ElGhazali}\ \emph {et~al.}(2017)\citenamefont {ElGhazali}, \citenamefont {Naumov}, \citenamefont {Mirhosseini}, \citenamefont {S{\"u}{\ss}}, \citenamefont {M{\"u}chler}, \citenamefont {Schnelle}, \citenamefont {Felser},\ and\ \citenamefont {Medvedev}}]{elghazali2017pressure}%
  \BibitemOpen
  \bibfield  {author} {\bibinfo {author} {\bibfnamefont {M.~A.}\ \bibnamefont {ElGhazali}}, \bibinfo {author} {\bibfnamefont {P.~G.}\ \bibnamefont {Naumov}}, \bibinfo {author} {\bibfnamefont {H.}~\bibnamefont {Mirhosseini}}, \bibinfo {author} {\bibfnamefont {V.}~\bibnamefont {S{\"u}{\ss}}}, \bibinfo {author} {\bibfnamefont {L.}~\bibnamefont {M{\"u}chler}}, \bibinfo {author} {\bibfnamefont {W.}~\bibnamefont {Schnelle}}, \bibinfo {author} {\bibfnamefont {C.}~\bibnamefont {Felser}},\ and\ \bibinfo {author} {\bibfnamefont {S.~A.}\ \bibnamefont {Medvedev}},\ }\bibfield  {title} {\bibinfo {title} {Pressure-induced superconductivity up to 13.1 k in the pyrite phase of palladium diselenide {PdSe}$_{2}$},\ }\href@noop {} {\bibfield  {journal} {\bibinfo  {journal} {Physical Review B}\ }\textbf {\bibinfo {volume} {96}},\ \bibinfo {pages} {060509} (\bibinfo {year} {2017})}\BibitemShut {NoStop}%
\bibitem [{\citenamefont {Nishiyama}\ \emph {et~al.}(2022{\natexlab{a}})\citenamefont {Nishiyama}, \citenamefont {Nishimura}, \citenamefont {Nishioka}, \citenamefont {Ueno}, \citenamefont {Iwamoto},\ and\ \citenamefont {Nagashio}}]{nishiyama2022bandgap}%
  \BibitemOpen
  \bibfield  {author} {\bibinfo {author} {\bibfnamefont {W.}~\bibnamefont {Nishiyama}}, \bibinfo {author} {\bibfnamefont {T.}~\bibnamefont {Nishimura}}, \bibinfo {author} {\bibfnamefont {M.}~\bibnamefont {Nishioka}}, \bibinfo {author} {\bibfnamefont {K.}~\bibnamefont {Ueno}}, \bibinfo {author} {\bibfnamefont {S.}~\bibnamefont {Iwamoto}},\ and\ \bibinfo {author} {\bibfnamefont {K.}~\bibnamefont {Nagashio}},\ }\bibfield  {title} {\bibinfo {title} {Is the bandgap of bulk {PdSe}$_{2}$ located truly in the far-infrared region? determination by fourier-transform photocurrent spectroscopy},\ }\href@noop {} {\bibfield  {journal} {\bibinfo  {journal} {Advanced Photonics Research}\ }\textbf {\bibinfo {volume} {3}},\ \bibinfo {pages} {2200231} (\bibinfo {year} {2022}{\natexlab{a}})}\BibitemShut {NoStop}%
\bibitem [{\citenamefont {Nishiyama}\ \emph {et~al.}(2022{\natexlab{b}})\citenamefont {Nishiyama}, \citenamefont {Nishimura}, \citenamefont {Ueno}, \citenamefont {Taniguchi}, \citenamefont {Watanabe},\ and\ \citenamefont {Nagashio}}]{nishiyama2022quantitative}%
  \BibitemOpen
  \bibfield  {author} {\bibinfo {author} {\bibfnamefont {W.}~\bibnamefont {Nishiyama}}, \bibinfo {author} {\bibfnamefont {T.}~\bibnamefont {Nishimura}}, \bibinfo {author} {\bibfnamefont {K.}~\bibnamefont {Ueno}}, \bibinfo {author} {\bibfnamefont {T.}~\bibnamefont {Taniguchi}}, \bibinfo {author} {\bibfnamefont {K.}~\bibnamefont {Watanabe}},\ and\ \bibinfo {author} {\bibfnamefont {K.}~\bibnamefont {Nagashio}},\ }\bibfield  {title} {\bibinfo {title} {Quantitative determination of contradictory bandgap values of bulk {PdSe}$_{2}$ from electrical transport properties},\ }\href@noop {} {\bibfield  {journal} {\bibinfo  {journal} {Advanced Functional Materials}\ }\textbf {\bibinfo {volume} {32}},\ \bibinfo {pages} {2108061} (\bibinfo {year} {2022}{\natexlab{b}})}\BibitemShut {NoStop}%
\bibitem [{\citenamefont {Sun}\ \emph {et~al.}(2015)\citenamefont {Sun}, \citenamefont {Shi}, \citenamefont {Siegrist},\ and\ \citenamefont {Singh}}]{sun2015electronic}%
  \BibitemOpen
  \bibfield  {author} {\bibinfo {author} {\bibfnamefont {J.}~\bibnamefont {Sun}}, \bibinfo {author} {\bibfnamefont {H.}~\bibnamefont {Shi}}, \bibinfo {author} {\bibfnamefont {T.}~\bibnamefont {Siegrist}},\ and\ \bibinfo {author} {\bibfnamefont {D.~J.}\ \bibnamefont {Singh}},\ }\bibfield  {title} {\bibinfo {title} {Electronic, transport, and optical properties of bulk and mono-layer {PdSe}$_{2}$},\ }\href@noop {} {\bibfield  {journal} {\bibinfo  {journal} {Applied Physics Letters}\ }\textbf {\bibinfo {volume} {107}} (\bibinfo {year} {2015})}\BibitemShut {NoStop}%
\bibitem [{\citenamefont {Wei}\ \emph {et~al.}(2022)\citenamefont {Wei}, \citenamefont {Lian}, \citenamefont {Zhang}, \citenamefont {Wang}, \citenamefont {Wang},\ and\ \citenamefont {Xu}}]{wei2022layer}%
  \BibitemOpen
  \bibfield  {author} {\bibinfo {author} {\bibfnamefont {M.}~\bibnamefont {Wei}}, \bibinfo {author} {\bibfnamefont {J.}~\bibnamefont {Lian}}, \bibinfo {author} {\bibfnamefont {Y.}~\bibnamefont {Zhang}}, \bibinfo {author} {\bibfnamefont {C.}~\bibnamefont {Wang}}, \bibinfo {author} {\bibfnamefont {Y.}~\bibnamefont {Wang}},\ and\ \bibinfo {author} {\bibfnamefont {Z.}~\bibnamefont {Xu}},\ }\bibfield  {title} {\bibinfo {title} {Layer-dependent optical and dielectric properties of centimeter-scale {PdSe}$_{2}$ films grown by chemical vapor deposition},\ }\href@noop {} {\bibfield  {journal} {\bibinfo  {journal} {npj 2\text{D} Materials and Applications}\ }\textbf {\bibinfo {volume} {6}},\ \bibinfo {pages} {1} (\bibinfo {year} {2022})}\BibitemShut {NoStop}%
\bibitem [{\citenamefont {Zhao}\ \emph {et~al.}(2020)\citenamefont {Zhao}, \citenamefont {Zhao}, \citenamefont {Zhao}, \citenamefont {Dai}, \citenamefont {Wei},\ and\ \citenamefont {Ma}}]{zhao2020electronic}%
  \BibitemOpen
  \bibfield  {author} {\bibinfo {author} {\bibfnamefont {X.}~\bibnamefont {Zhao}}, \bibinfo {author} {\bibfnamefont {Q.}~\bibnamefont {Zhao}}, \bibinfo {author} {\bibfnamefont {B.}~\bibnamefont {Zhao}}, \bibinfo {author} {\bibfnamefont {X.}~\bibnamefont {Dai}}, \bibinfo {author} {\bibfnamefont {S.}~\bibnamefont {Wei}},\ and\ \bibinfo {author} {\bibfnamefont {Y.}~\bibnamefont {Ma}},\ }\bibfield  {title} {\bibinfo {title} {Electronic and optical properties of {PdSe}$_{2}$ from monolayer to trilayer},\ }\href@noop {} {\bibfield  {journal} {\bibinfo  {journal} {Superlattices and Microstructures}\ }\textbf {\bibinfo {volume} {142}},\ \bibinfo {pages} {106514} (\bibinfo {year} {2020})}\BibitemShut {NoStop}%
\bibitem [{\citenamefont {Kuklin}\ and\ \citenamefont {{\AA}gren}(2019)}]{kuklin2019quasiparticle}%
  \BibitemOpen
  \bibfield  {author} {\bibinfo {author} {\bibfnamefont {A.~V.}\ \bibnamefont {Kuklin}}\ and\ \bibinfo {author} {\bibfnamefont {H.}~\bibnamefont {{\AA}gren}},\ }\bibfield  {title} {\bibinfo {title} {Quasiparticle electronic structure and optical spectra of single-layer and bilayer {PdSe}$_{2}$: Proximity and defect-induced band gap renormalization},\ }\href@noop {} {\bibfield  {journal} {\bibinfo  {journal} {Physical Review B}\ }\textbf {\bibinfo {volume} {99}},\ \bibinfo {pages} {245114} (\bibinfo {year} {2019})}\BibitemShut {NoStop}%
\bibitem [{\citenamefont {Li}\ \emph {et~al.}(2021)\citenamefont {Li}, \citenamefont {Peng}, \citenamefont {Lin}, \citenamefont {Leng}, \citenamefont {Zhang}, \citenamefont {Pang}, \citenamefont {Tan}, \citenamefont {Monserrat},\ and\ \citenamefont {Chen}}]{li2021phonon}%
  \BibitemOpen
  \bibfield  {author} {\bibinfo {author} {\bibfnamefont {Z.}~\bibnamefont {Li}}, \bibinfo {author} {\bibfnamefont {B.}~\bibnamefont {Peng}}, \bibinfo {author} {\bibfnamefont {M.-L.}\ \bibnamefont {Lin}}, \bibinfo {author} {\bibfnamefont {Y.-C.}\ \bibnamefont {Leng}}, \bibinfo {author} {\bibfnamefont {B.}~\bibnamefont {Zhang}}, \bibinfo {author} {\bibfnamefont {C.}~\bibnamefont {Pang}}, \bibinfo {author} {\bibfnamefont {P.-H.}\ \bibnamefont {Tan}}, \bibinfo {author} {\bibfnamefont {B.}~\bibnamefont {Monserrat}},\ and\ \bibinfo {author} {\bibfnamefont {F.}~\bibnamefont {Chen}},\ }\bibfield  {title} {\bibinfo {title} {Phonon-assisted electronic states modulation of few-layer {PdSe}$_{2}$ at terahertz frequencies},\ }\href@noop {} {\bibfield  {journal} {\bibinfo  {journal} {npj 2\text{D} Materials and Applications}\ }\textbf {\bibinfo {volume} {5}},\ \bibinfo {pages} {87} (\bibinfo {year} {2021})}\BibitemShut {NoStop}%
\bibitem [{\citenamefont {Cattelan}\ \emph {et~al.}(2021)\citenamefont {Cattelan}, \citenamefont {Sayers}, \citenamefont {Wolverson},\ and\ \citenamefont {Carpene}}]{cattelan2021site}%
  \BibitemOpen
  \bibfield  {author} {\bibinfo {author} {\bibfnamefont {M.}~\bibnamefont {Cattelan}}, \bibinfo {author} {\bibfnamefont {C.}~\bibnamefont {Sayers}}, \bibinfo {author} {\bibfnamefont {D.}~\bibnamefont {Wolverson}},\ and\ \bibinfo {author} {\bibfnamefont {E.}~\bibnamefont {Carpene}},\ }\bibfield  {title} {\bibinfo {title} {Site-specific symmetry sensitivity of angle-resolved photoemission spectroscopy in layered palladium diselenide},\ }\href@noop {} {\bibfield  {journal} {\bibinfo  {journal} {2\text{D} Materials}\ }\textbf {\bibinfo {volume} {8}},\ \bibinfo {pages} {045036} (\bibinfo {year} {2021})}\BibitemShut {NoStop}%
\bibitem [{\citenamefont {Gu}\ \emph {et~al.}(2022)\citenamefont {Gu}, \citenamefont {Liu}, \citenamefont {Chen}, \citenamefont {Liang}, \citenamefont {Guo}, \citenamefont {Yang}, \citenamefont {Zhou}, \citenamefont {Jozwiak}, \citenamefont {Bostwick}, \citenamefont {Liu} \emph {et~al.}}]{gu2022low}%
  \BibitemOpen
  \bibfield  {author} {\bibinfo {author} {\bibfnamefont {C.}~\bibnamefont {Gu}}, \bibinfo {author} {\bibfnamefont {X.}~\bibnamefont {Liu}}, \bibinfo {author} {\bibfnamefont {C.}~\bibnamefont {Chen}}, \bibinfo {author} {\bibfnamefont {A.}~\bibnamefont {Liang}}, \bibinfo {author} {\bibfnamefont {W.}~\bibnamefont {Guo}}, \bibinfo {author} {\bibfnamefont {X.}~\bibnamefont {Yang}}, \bibinfo {author} {\bibfnamefont {J.}~\bibnamefont {Zhou}}, \bibinfo {author} {\bibfnamefont {C.}~\bibnamefont {Jozwiak}}, \bibinfo {author} {\bibfnamefont {A.}~\bibnamefont {Bostwick}}, \bibinfo {author} {\bibfnamefont {Z.}~\bibnamefont {Liu}}, \emph {et~al.},\ }\bibfield  {title} {\bibinfo {title} {Low-lying electronic states with giant linear dichroic ratio observed in {PdSe}$_{2}$},\ }\href@noop {} {\bibfield  {journal} {\bibinfo  {journal} {Physical Review B}\ }\textbf {\bibinfo {volume} {106}},\ \bibinfo {pages} {L121110} (\bibinfo {year} {2022})}\BibitemShut {NoStop}%
\bibitem [{\citenamefont {Chung}\ \emph {et~al.}(2024)\citenamefont {Chung}, \citenamefont {Kim}, \citenamefont {Kim}, \citenamefont {Cha}, \citenamefont {Park}, \citenamefont {Park}, \citenamefont {Yi}, \citenamefont {Song}, \citenamefont {Ryu}, \citenamefont {Lee} \emph {et~al.}}]{chung2024dark}%
  \BibitemOpen
  \bibfield  {author} {\bibinfo {author} {\bibfnamefont {Y.}~\bibnamefont {Chung}}, \bibinfo {author} {\bibfnamefont {M.}~\bibnamefont {Kim}}, \bibinfo {author} {\bibfnamefont {Y.}~\bibnamefont {Kim}}, \bibinfo {author} {\bibfnamefont {S.}~\bibnamefont {Cha}}, \bibinfo {author} {\bibfnamefont {J.~W.}\ \bibnamefont {Park}}, \bibinfo {author} {\bibfnamefont {J.}~\bibnamefont {Park}}, \bibinfo {author} {\bibfnamefont {Y.}~\bibnamefont {Yi}}, \bibinfo {author} {\bibfnamefont {D.}~\bibnamefont {Song}}, \bibinfo {author} {\bibfnamefont {J.~H.}\ \bibnamefont {Ryu}}, \bibinfo {author} {\bibfnamefont {K.}~\bibnamefont {Lee}}, \emph {et~al.},\ }\bibfield  {title} {\bibinfo {title} {Dark states of electrons in a quantum system with two pairs of sublattices},\ }\href@noop {} {\bibfield  {journal} {\bibinfo  {journal} {Nature Physics}\ }\textbf {\bibinfo {volume} {20}},\ \bibinfo {pages} {1582} (\bibinfo {year} {2024})}\BibitemShut {NoStop}%
\bibitem [{\citenamefont {Ryu}\ \emph {et~al.}(2022)\citenamefont {Ryu}, \citenamefont {Kim}, \citenamefont {Kim}, \citenamefont {Kim}, \citenamefont {Kim}, \citenamefont {Park}, \citenamefont {Park}, \citenamefont {Kim}, \citenamefont {Ko},\ and\ \citenamefont {Lee}}]{ryu2022direct}%
  \BibitemOpen
  \bibfield  {author} {\bibinfo {author} {\bibfnamefont {J.~H.}\ \bibnamefont {Ryu}}, \bibinfo {author} {\bibfnamefont {J.-G.}\ \bibnamefont {Kim}}, \bibinfo {author} {\bibfnamefont {B.}~\bibnamefont {Kim}}, \bibinfo {author} {\bibfnamefont {K.}~\bibnamefont {Kim}}, \bibinfo {author} {\bibfnamefont {S.}~\bibnamefont {Kim}}, \bibinfo {author} {\bibfnamefont {J.-H.}\ \bibnamefont {Park}}, \bibinfo {author} {\bibfnamefont {B.-G.}\ \bibnamefont {Park}}, \bibinfo {author} {\bibfnamefont {Y.}~\bibnamefont {Kim}}, \bibinfo {author} {\bibfnamefont {K.-T.}\ \bibnamefont {Ko}},\ and\ \bibinfo {author} {\bibfnamefont {K.}~\bibnamefont {Lee}},\ }\bibfield  {title} {\bibinfo {title} {Direct observation of orbital driven strong interlayer coupling in puckered two-dimensional {PdSe}$_{2}$},\ }\href@noop {} {\bibfield  {journal} {\bibinfo  {journal} {Small}\ }\textbf {\bibinfo {volume} {18}},\ \bibinfo {pages} {2106053} (\bibinfo {year} {2022})}\BibitemShut {NoStop}%
\bibitem [{\citenamefont {Kim}\ \emph {et~al.}(2015)\citenamefont {Kim}, \citenamefont {Baik}, \citenamefont {Ryu}, \citenamefont {Sohn}, \citenamefont {Park}, \citenamefont {Park}, \citenamefont {Denlinger}, \citenamefont {Yi}, \citenamefont {Choi},\ and\ \citenamefont {Kim}}]{kim2015observation}%
  \BibitemOpen
  \bibfield  {author} {\bibinfo {author} {\bibfnamefont {J.}~\bibnamefont {Kim}}, \bibinfo {author} {\bibfnamefont {S.~S.}\ \bibnamefont {Baik}}, \bibinfo {author} {\bibfnamefont {S.~H.}\ \bibnamefont {Ryu}}, \bibinfo {author} {\bibfnamefont {Y.}~\bibnamefont {Sohn}}, \bibinfo {author} {\bibfnamefont {S.}~\bibnamefont {Park}}, \bibinfo {author} {\bibfnamefont {B.-G.}\ \bibnamefont {Park}}, \bibinfo {author} {\bibfnamefont {J.}~\bibnamefont {Denlinger}}, \bibinfo {author} {\bibfnamefont {Y.}~\bibnamefont {Yi}}, \bibinfo {author} {\bibfnamefont {H.~J.}\ \bibnamefont {Choi}},\ and\ \bibinfo {author} {\bibfnamefont {K.~S.}\ \bibnamefont {Kim}},\ }\bibfield  {title} {\bibinfo {title} {Observation of tunable band gap and anisotropic {Dirac} semimetal state in black phosphorus},\ }\href@noop {} {\bibfield  {journal} {\bibinfo  {journal} {Science}\ }\textbf {\bibinfo {volume} {349}},\ \bibinfo {pages} {723} (\bibinfo {year} {2015})}\BibitemShut {NoStop}%
\bibitem [{\citenamefont {Shan}\ \emph {et~al.}(2025)\citenamefont {Shan}, \citenamefont {Meng}, \citenamefont {Li}, \citenamefont {Liu}, \citenamefont {Liu}, \citenamefont {Sun}, \citenamefont {Dong},\ and\ \citenamefont {Chen}}]{shan2025matrix}%
  \BibitemOpen
  \bibfield  {author} {\bibinfo {author} {\bibfnamefont {S.}~\bibnamefont {Shan}}, \bibinfo {author} {\bibfnamefont {F.}~\bibnamefont {Meng}}, \bibinfo {author} {\bibfnamefont {T.}~\bibnamefont {Li}}, \bibinfo {author} {\bibfnamefont {Z.}~\bibnamefont {Liu}}, \bibinfo {author} {\bibfnamefont {Y.}~\bibnamefont {Liu}}, \bibinfo {author} {\bibfnamefont {Z.}~\bibnamefont {Sun}}, \bibinfo {author} {\bibfnamefont {J.}~\bibnamefont {Dong}},\ and\ \bibinfo {author} {\bibfnamefont {Z.}~\bibnamefont {Chen}},\ }\bibfield  {title} {\bibinfo {title} {Matrix element effect in layered {PdSe}$_{2}$},\ }\href@noop {} {\bibfield  {journal} {\bibinfo  {journal} {The Journal of Physical Chemistry C}\ } (\bibinfo {year} {2025})}\BibitemShut {NoStop}%
\bibitem [{\citenamefont {Boschini}\ \emph {et~al.}(2024)\citenamefont {Boschini}, \citenamefont {Zonno},\ and\ \citenamefont {Damascelli}}]{boschini2024time}%
  \BibitemOpen
  \bibfield  {author} {\bibinfo {author} {\bibfnamefont {F.}~\bibnamefont {Boschini}}, \bibinfo {author} {\bibfnamefont {M.}~\bibnamefont {Zonno}},\ and\ \bibinfo {author} {\bibfnamefont {A.}~\bibnamefont {Damascelli}},\ }\bibfield  {title} {\bibinfo {title} {Time-resolved {ARPES} studies of quantum materials},\ }\href@noop {} {\bibfield  {journal} {\bibinfo  {journal} {Reviews of Modern Physics}\ }\textbf {\bibinfo {volume} {96}},\ \bibinfo {pages} {015003} (\bibinfo {year} {2024})}\BibitemShut {NoStop}%
\bibitem [{\citenamefont {Aeschlimann}\ \emph {et~al.}(2025)\citenamefont {Aeschlimann}, \citenamefont {Bange}, \citenamefont {Bauer}, \citenamefont {Bovensiepen}, \citenamefont {Elmers}, \citenamefont {Fauster}, \citenamefont {Gierster}, \citenamefont {H{\"o}fer}, \citenamefont {Huber}, \citenamefont {Li} \emph {et~al.}}]{aeschlimann2025time}%
  \BibitemOpen
  \bibfield  {author} {\bibinfo {author} {\bibfnamefont {M.}~\bibnamefont {Aeschlimann}}, \bibinfo {author} {\bibfnamefont {J.~P.}\ \bibnamefont {Bange}}, \bibinfo {author} {\bibfnamefont {M.}~\bibnamefont {Bauer}}, \bibinfo {author} {\bibfnamefont {U.}~\bibnamefont {Bovensiepen}}, \bibinfo {author} {\bibfnamefont {H.-J.}\ \bibnamefont {Elmers}}, \bibinfo {author} {\bibfnamefont {T.}~\bibnamefont {Fauster}}, \bibinfo {author} {\bibfnamefont {L.}~\bibnamefont {Gierster}}, \bibinfo {author} {\bibfnamefont {U.}~\bibnamefont {H{\"o}fer}}, \bibinfo {author} {\bibfnamefont {R.}~\bibnamefont {Huber}}, \bibinfo {author} {\bibfnamefont {A.}~\bibnamefont {Li}}, \emph {et~al.},\ }\bibfield  {title} {\bibinfo {title} {Time-resolved photoelectron spectroscopy at surfaces},\ }\href@noop {} {\bibfield  {journal} {\bibinfo  {journal} {Surface Science}\ }\textbf {\bibinfo {volume} {753}},\ \bibinfo {pages} {122631} (\bibinfo {year} {2025})}\BibitemShut {NoStop}%
\bibitem [{\citenamefont {Hedayat}\ \emph {et~al.}(2021{\natexlab{a}})\citenamefont {Hedayat}, \citenamefont {Bugini}, \citenamefont {Yi}, \citenamefont {Chen}, \citenamefont {Zhou}, \citenamefont {Cerullo}, \citenamefont {Dallera},\ and\ \citenamefont {Carpene}}]{hedayat2021ultrafast}%
  \BibitemOpen
  \bibfield  {author} {\bibinfo {author} {\bibfnamefont {H.}~\bibnamefont {Hedayat}}, \bibinfo {author} {\bibfnamefont {D.}~\bibnamefont {Bugini}}, \bibinfo {author} {\bibfnamefont {H.}~\bibnamefont {Yi}}, \bibinfo {author} {\bibfnamefont {C.}~\bibnamefont {Chen}}, \bibinfo {author} {\bibfnamefont {X.}~\bibnamefont {Zhou}}, \bibinfo {author} {\bibfnamefont {G.}~\bibnamefont {Cerullo}}, \bibinfo {author} {\bibfnamefont {C.}~\bibnamefont {Dallera}},\ and\ \bibinfo {author} {\bibfnamefont {E.}~\bibnamefont {Carpene}},\ }\bibfield  {title} {\bibinfo {title} {Ultrafast evolution of bulk, surface and surface resonance states in photoexcited {Bi}$_{2}${Te}$_{3}$},\ }\href@noop {} {\bibfield  {journal} {\bibinfo  {journal} {Scientific Reports}\ }\textbf {\bibinfo {volume} {11}},\ \bibinfo {pages} {4924} (\bibinfo {year} {2021}{\natexlab{a}})}\BibitemShut {NoStop}%
\bibitem [{\citenamefont {Hedayat}\ \emph {et~al.}(2021{\natexlab{b}})\citenamefont {Hedayat}, \citenamefont {Ceraso}, \citenamefont {Soavi}, \citenamefont {Akhavan}, \citenamefont {Cadore}, \citenamefont {Dallera}, \citenamefont {Cerullo}, \citenamefont {Ferrari},\ and\ \citenamefont {Carpene}}]{hedayat2021non}%
  \BibitemOpen
  \bibfield  {author} {\bibinfo {author} {\bibfnamefont {H.}~\bibnamefont {Hedayat}}, \bibinfo {author} {\bibfnamefont {A.}~\bibnamefont {Ceraso}}, \bibinfo {author} {\bibfnamefont {G.}~\bibnamefont {Soavi}}, \bibinfo {author} {\bibfnamefont {S.}~\bibnamefont {Akhavan}}, \bibinfo {author} {\bibfnamefont {A.}~\bibnamefont {Cadore}}, \bibinfo {author} {\bibfnamefont {C.}~\bibnamefont {Dallera}}, \bibinfo {author} {\bibfnamefont {G.}~\bibnamefont {Cerullo}}, \bibinfo {author} {\bibfnamefont {A.}~\bibnamefont {Ferrari}},\ and\ \bibinfo {author} {\bibfnamefont {E.}~\bibnamefont {Carpene}},\ }\bibfield  {title} {\bibinfo {title} {Non-equilibrium band broadening, gap renormalization and band inversion in black phosphorus},\ }\href@noop {} {\bibfield  {journal} {\bibinfo  {journal} {2\text{D} Materials}\ }\textbf {\bibinfo {volume} {8}},\ \bibinfo {pages} {025020} (\bibinfo {year} {2021}{\natexlab{b}})}\BibitemShut {NoStop}%
\bibitem [{\citenamefont {Kremer}\ \emph {et~al.}(2021)\citenamefont {Kremer}, \citenamefont {Rumo}, \citenamefont {Yue}, \citenamefont {Pulkkinen}, \citenamefont {Nicholson}, \citenamefont {Jaouen}, \citenamefont {von Rohr}, \citenamefont {Werner},\ and\ \citenamefont {Monney}}]{kremer2021ultrafast}%
  \BibitemOpen
  \bibfield  {author} {\bibinfo {author} {\bibfnamefont {G.}~\bibnamefont {Kremer}}, \bibinfo {author} {\bibfnamefont {M.}~\bibnamefont {Rumo}}, \bibinfo {author} {\bibfnamefont {C.}~\bibnamefont {Yue}}, \bibinfo {author} {\bibfnamefont {A.}~\bibnamefont {Pulkkinen}}, \bibinfo {author} {\bibfnamefont {C.~W.}\ \bibnamefont {Nicholson}}, \bibinfo {author} {\bibfnamefont {T.}~\bibnamefont {Jaouen}}, \bibinfo {author} {\bibfnamefont {F.~O.}\ \bibnamefont {von Rohr}}, \bibinfo {author} {\bibfnamefont {P.}~\bibnamefont {Werner}},\ and\ \bibinfo {author} {\bibfnamefont {C.}~\bibnamefont {Monney}},\ }\bibfield  {title} {\bibinfo {title} {Ultrafast dynamics of the surface photovoltage in potassium-doped black phosphorus},\ }\href@noop {} {\bibfield  {journal} {\bibinfo  {journal} {Physical Review B}\ }\textbf {\bibinfo {volume} {104}},\ \bibinfo {pages} {035125} (\bibinfo {year} {2021})}\BibitemShut {NoStop}%
\bibitem [{\citenamefont {Chen}\ \emph {et~al.}(2018)\citenamefont {Chen}, \citenamefont {Dong}, \citenamefont {Papalazarou}, \citenamefont {Marsi}, \citenamefont {Giorgetti}, \citenamefont {Zhang}, \citenamefont {Tian}, \citenamefont {Rueff}, \citenamefont {Taleb-Ibrahimi},\ and\ \citenamefont {Perfetti}}]{chen2018band}%
  \BibitemOpen
  \bibfield  {author} {\bibinfo {author} {\bibfnamefont {Z.}~\bibnamefont {Chen}}, \bibinfo {author} {\bibfnamefont {J.}~\bibnamefont {Dong}}, \bibinfo {author} {\bibfnamefont {E.}~\bibnamefont {Papalazarou}}, \bibinfo {author} {\bibfnamefont {M.}~\bibnamefont {Marsi}}, \bibinfo {author} {\bibfnamefont {C.}~\bibnamefont {Giorgetti}}, \bibinfo {author} {\bibfnamefont {Z.}~\bibnamefont {Zhang}}, \bibinfo {author} {\bibfnamefont {B.}~\bibnamefont {Tian}}, \bibinfo {author} {\bibfnamefont {J.-P.}\ \bibnamefont {Rueff}}, \bibinfo {author} {\bibfnamefont {A.}~\bibnamefont {Taleb-Ibrahimi}},\ and\ \bibinfo {author} {\bibfnamefont {L.}~\bibnamefont {Perfetti}},\ }\bibfield  {title} {\bibinfo {title} {Band gap renormalization, carrier multiplication, and stark broadening in photoexcited black phosphorus},\ }\href@noop {} {\bibfield  {journal} {\bibinfo  {journal} {Nano letters}\ }\textbf {\bibinfo {volume} {19}},\ \bibinfo {pages} {488} (\bibinfo {year} {2018})}\BibitemShut {NoStop}%
\bibitem [{\citenamefont {Roth}\ \emph {et~al.}(2019)\citenamefont {Roth}, \citenamefont {Crepaldi}, \citenamefont {Puppin}, \citenamefont {Gatti}, \citenamefont {Bugini}, \citenamefont {Grimaldi}, \citenamefont {Barrilot}, \citenamefont {Arrell}, \citenamefont {Frassetto}, \citenamefont {Poletto} \emph {et~al.}}]{roth2019photocarrier}%
  \BibitemOpen
  \bibfield  {author} {\bibinfo {author} {\bibfnamefont {S.}~\bibnamefont {Roth}}, \bibinfo {author} {\bibfnamefont {A.}~\bibnamefont {Crepaldi}}, \bibinfo {author} {\bibfnamefont {M.}~\bibnamefont {Puppin}}, \bibinfo {author} {\bibfnamefont {G.}~\bibnamefont {Gatti}}, \bibinfo {author} {\bibfnamefont {D.}~\bibnamefont {Bugini}}, \bibinfo {author} {\bibfnamefont {I.}~\bibnamefont {Grimaldi}}, \bibinfo {author} {\bibfnamefont {T.}~\bibnamefont {Barrilot}}, \bibinfo {author} {\bibfnamefont {C.}~\bibnamefont {Arrell}}, \bibinfo {author} {\bibfnamefont {F.}~\bibnamefont {Frassetto}}, \bibinfo {author} {\bibfnamefont {L.}~\bibnamefont {Poletto}}, \emph {et~al.},\ }\bibfield  {title} {\bibinfo {title} {Photocarrier-induced band-gap renormalization and ultrafast charge dynamics in black phosphorus},\ }\href@noop {} {\bibfield  {journal} {\bibinfo  {journal} {2\text{D} Materials}\ }\textbf {\bibinfo {volume} {6}},\ \bibinfo {pages} {031001} (\bibinfo {year} {2019})}\BibitemShut {NoStop}%
\bibitem [{\citenamefont {Gierster}\ \emph {et~al.}(2021)\citenamefont {Gierster}, \citenamefont {Vempati},\ and\ \citenamefont {St{\"a}hler}}]{gierster2021ultrafast}%
  \BibitemOpen
  \bibfield  {author} {\bibinfo {author} {\bibfnamefont {L.}~\bibnamefont {Gierster}}, \bibinfo {author} {\bibfnamefont {S.}~\bibnamefont {Vempati}},\ and\ \bibinfo {author} {\bibfnamefont {J.}~\bibnamefont {St{\"a}hler}},\ }\bibfield  {title} {\bibinfo {title} {Ultrafast generation and decay of a surface metal},\ }\href@noop {} {\bibfield  {journal} {\bibinfo  {journal} {Nature Communications}\ }\textbf {\bibinfo {volume} {12}},\ \bibinfo {pages} {978} (\bibinfo {year} {2021})}\BibitemShut {NoStop}%
\bibitem [{\citenamefont {Weinelt}\ \emph {et~al.}(2005)\citenamefont {Weinelt}, \citenamefont {Kutschera}, \citenamefont {Schmidt}, \citenamefont {Orth}, \citenamefont {Fauster},\ and\ \citenamefont {Rohlfing}}]{weinelt2005electronic}%
  \BibitemOpen
  \bibfield  {author} {\bibinfo {author} {\bibfnamefont {M.}~\bibnamefont {Weinelt}}, \bibinfo {author} {\bibfnamefont {M.}~\bibnamefont {Kutschera}}, \bibinfo {author} {\bibfnamefont {R.}~\bibnamefont {Schmidt}}, \bibinfo {author} {\bibfnamefont {C.}~\bibnamefont {Orth}}, \bibinfo {author} {\bibfnamefont {T.}~\bibnamefont {Fauster}},\ and\ \bibinfo {author} {\bibfnamefont {M.}~\bibnamefont {Rohlfing}},\ }\bibfield  {title} {\bibinfo {title} {Electronic structure and electron dynamics at {Si} (100)},\ }\href@noop {} {\bibfield  {journal} {\bibinfo  {journal} {Applied Physics A}\ }\textbf {\bibinfo {volume} {80}},\ \bibinfo {pages} {995} (\bibinfo {year} {2005})}\BibitemShut {NoStop}%
\bibitem [{\citenamefont {Weinelt}\ \emph {et~al.}(2004)\citenamefont {Weinelt}, \citenamefont {Kutschera}, \citenamefont {Fauster},\ and\ \citenamefont {Rohlfing}}]{weinelt2004dynamics}%
  \BibitemOpen
  \bibfield  {author} {\bibinfo {author} {\bibfnamefont {M.}~\bibnamefont {Weinelt}}, \bibinfo {author} {\bibfnamefont {M.}~\bibnamefont {Kutschera}}, \bibinfo {author} {\bibfnamefont {T.}~\bibnamefont {Fauster}},\ and\ \bibinfo {author} {\bibfnamefont {M.}~\bibnamefont {Rohlfing}},\ }\bibfield  {title} {\bibinfo {title} {Dynamics of exciton formation at the {Si} (100) c (4$\times$ 2) surface},\ }\href@noop {} {\bibfield  {journal} {\bibinfo  {journal} {Physical review letters}\ }\textbf {\bibinfo {volume} {92}},\ \bibinfo {pages} {126801} (\bibinfo {year} {2004})}\BibitemShut {NoStop}%
\bibitem [{\citenamefont {Ciocys}\ \emph {et~al.}(2019)\citenamefont {Ciocys}, \citenamefont {Morimoto}, \citenamefont {Moore},\ and\ \citenamefont {Lanzara}}]{ciocys2019tracking}%
  \BibitemOpen
  \bibfield  {author} {\bibinfo {author} {\bibfnamefont {S.}~\bibnamefont {Ciocys}}, \bibinfo {author} {\bibfnamefont {T.}~\bibnamefont {Morimoto}}, \bibinfo {author} {\bibfnamefont {J.}~\bibnamefont {Moore}},\ and\ \bibinfo {author} {\bibfnamefont {A.}~\bibnamefont {Lanzara}},\ }\bibfield  {title} {\bibinfo {title} {Tracking surface photovoltage dipole geometry in {Bi}$_2${Se}$_3$ with time-resolved photoemission},\ }\href@noop {} {\bibfield  {journal} {\bibinfo  {journal} {Journal of Statistical Mechanics: Theory and Experiment}\ }\textbf {\bibinfo {volume} {2019}},\ \bibinfo {pages} {104008} (\bibinfo {year} {2019})}\BibitemShut {NoStop}%
\bibitem [{\citenamefont {Yang}\ \emph {et~al.}(2014)\citenamefont {Yang}, \citenamefont {Sobota}, \citenamefont {Kirchmann},\ and\ \citenamefont {Shen}}]{yang2014electron}%
  \BibitemOpen
  \bibfield  {author} {\bibinfo {author} {\bibfnamefont {S.-L.}\ \bibnamefont {Yang}}, \bibinfo {author} {\bibfnamefont {J.~A.}\ \bibnamefont {Sobota}}, \bibinfo {author} {\bibfnamefont {P.~S.}\ \bibnamefont {Kirchmann}},\ and\ \bibinfo {author} {\bibfnamefont {Z.-X.}\ \bibnamefont {Shen}},\ }\bibfield  {title} {\bibinfo {title} {Electron propagation from a photo-excited surface: implications for time-resolved photoemission},\ }\href@noop {} {\bibfield  {journal} {\bibinfo  {journal} {Applied Physics A}\ }\textbf {\bibinfo {volume} {116}},\ \bibinfo {pages} {85} (\bibinfo {year} {2014})}\BibitemShut {NoStop}%
\bibitem [{\citenamefont {Ulstrup}\ \emph {et~al.}(2015)\citenamefont {Ulstrup}, \citenamefont {Johannsen}, \citenamefont {Cilento}, \citenamefont {Crepaldi}, \citenamefont {Miwa}, \citenamefont {Zacchigna}, \citenamefont {Cacho}, \citenamefont {Chapman}, \citenamefont {Springate}, \citenamefont {Fromm} \emph {et~al.}}]{ulstrup2015ramifications}%
  \BibitemOpen
  \bibfield  {author} {\bibinfo {author} {\bibfnamefont {S.}~\bibnamefont {Ulstrup}}, \bibinfo {author} {\bibfnamefont {J.~C.}\ \bibnamefont {Johannsen}}, \bibinfo {author} {\bibfnamefont {F.}~\bibnamefont {Cilento}}, \bibinfo {author} {\bibfnamefont {A.}~\bibnamefont {Crepaldi}}, \bibinfo {author} {\bibfnamefont {J.~A.}\ \bibnamefont {Miwa}}, \bibinfo {author} {\bibfnamefont {M.}~\bibnamefont {Zacchigna}}, \bibinfo {author} {\bibfnamefont {C.}~\bibnamefont {Cacho}}, \bibinfo {author} {\bibfnamefont {R.~T.}\ \bibnamefont {Chapman}}, \bibinfo {author} {\bibfnamefont {E.}~\bibnamefont {Springate}}, \bibinfo {author} {\bibfnamefont {F.}~\bibnamefont {Fromm}}, \emph {et~al.},\ }\bibfield  {title} {\bibinfo {title} {Ramifications of optical pumping on the interpretation of time-resolved photoemission experiments on graphene},\ }\href@noop {} {\bibfield  {journal} {\bibinfo  {journal} {Journal of Electron Spectroscopy and Related Phenomena}\ }\textbf {\bibinfo {volume} {200}},\ \bibinfo {pages} {340} (\bibinfo
  {year} {2015})}\BibitemShut {NoStop}%
\bibitem [{\citenamefont {Ciocys}\ \emph {et~al.}(2020)\citenamefont {Ciocys}, \citenamefont {Morimoto}, \citenamefont {Mori}, \citenamefont {Gotlieb}, \citenamefont {Hussain}, \citenamefont {Analytis}, \citenamefont {Moore},\ and\ \citenamefont {Lanzara}}]{ciocys2020manipulating}%
  \BibitemOpen
  \bibfield  {author} {\bibinfo {author} {\bibfnamefont {S.}~\bibnamefont {Ciocys}}, \bibinfo {author} {\bibfnamefont {T.}~\bibnamefont {Morimoto}}, \bibinfo {author} {\bibfnamefont {R.}~\bibnamefont {Mori}}, \bibinfo {author} {\bibfnamefont {K.}~\bibnamefont {Gotlieb}}, \bibinfo {author} {\bibfnamefont {Z.}~\bibnamefont {Hussain}}, \bibinfo {author} {\bibfnamefont {J.~G.}\ \bibnamefont {Analytis}}, \bibinfo {author} {\bibfnamefont {J.~E.}\ \bibnamefont {Moore}},\ and\ \bibinfo {author} {\bibfnamefont {A.}~\bibnamefont {Lanzara}},\ }\bibfield  {title} {\bibinfo {title} {Manipulating long-lived topological surface photovoltage in bulk-insulating topological insulators ={Bi}$_{2}${Se}$_{3}$ and {Bi}$_{2}${Te}$_{3}$},\ }\href@noop {} {\bibfield  {journal} {\bibinfo  {journal} {npj Quantum Materials}\ }\textbf {\bibinfo {volume} {5}},\ \bibinfo {pages} {1} (\bibinfo {year} {2020})}\BibitemShut {NoStop}%
\bibitem [{\citenamefont {Kuklin}\ \emph {et~al.}(2021)\citenamefont {Kuklin}, \citenamefont {Begunovich}, \citenamefont {Gao}, \citenamefont {Zhang},\ and\ \citenamefont {{\AA}gren}}]{kuklin2021point}%
  \BibitemOpen
  \bibfield  {author} {\bibinfo {author} {\bibfnamefont {A.~V.}\ \bibnamefont {Kuklin}}, \bibinfo {author} {\bibfnamefont {L.~V.}\ \bibnamefont {Begunovich}}, \bibinfo {author} {\bibfnamefont {L.}~\bibnamefont {Gao}}, \bibinfo {author} {\bibfnamefont {H.}~\bibnamefont {Zhang}},\ and\ \bibinfo {author} {\bibfnamefont {H.}~\bibnamefont {{\AA}gren}},\ }\bibfield  {title} {\bibinfo {title} {Point and complex defects in monolayer {PdSe}$_{2}$: Evolution of electronic structure and emergence of magnetism},\ }\href@noop {} {\bibfield  {journal} {\bibinfo  {journal} {Physical Review B}\ }\textbf {\bibinfo {volume} {104}},\ \bibinfo {pages} {134109} (\bibinfo {year} {2021})}\BibitemShut {NoStop}%
\bibitem [{\citenamefont {Jena}\ \emph {et~al.}(2023)\citenamefont {Jena}, \citenamefont {Hossain}, \citenamefont {Nath}, \citenamefont {Sarma}, \citenamefont {Sugimoto}, \citenamefont {Fujii},\ and\ \citenamefont {Giri}}]{jena2023evidence}%
  \BibitemOpen
  \bibfield  {author} {\bibinfo {author} {\bibfnamefont {T.}~\bibnamefont {Jena}}, \bibinfo {author} {\bibfnamefont {M.~T.}\ \bibnamefont {Hossain}}, \bibinfo {author} {\bibfnamefont {U.}~\bibnamefont {Nath}}, \bibinfo {author} {\bibfnamefont {M.}~\bibnamefont {Sarma}}, \bibinfo {author} {\bibfnamefont {H.}~\bibnamefont {Sugimoto}}, \bibinfo {author} {\bibfnamefont {M.}~\bibnamefont {Fujii}},\ and\ \bibinfo {author} {\bibfnamefont {P.}~\bibnamefont {Giri}},\ }\bibfield  {title} {\bibinfo {title} {Evidence for intrinsic defects and nanopores as hotspots in 2\text{D} \text{PdSe}$_{2}$ dendrites for plasmon-free sers substrate with a high enhancement factor},\ }\href@noop {} {\bibfield  {journal} {\bibinfo  {journal} {npj 2\text{D} Materials and Applications}\ }\textbf {\bibinfo {volume} {7}},\ \bibinfo {pages} {8} (\bibinfo {year} {2023})}\BibitemShut {NoStop}%
\bibitem [{\citenamefont {Nguyen}\ \emph {et~al.}(2018)\citenamefont {Nguyen}, \citenamefont {Liang}, \citenamefont {Zou}, \citenamefont {Fu}, \citenamefont {Oyedele}, \citenamefont {Sumpter}, \citenamefont {Liu}, \citenamefont {Gai}, \citenamefont {Xiao},\ and\ \citenamefont {Li}}]{nguyen20183d}%
  \BibitemOpen
  \bibfield  {author} {\bibinfo {author} {\bibfnamefont {G.~D.}\ \bibnamefont {Nguyen}}, \bibinfo {author} {\bibfnamefont {L.}~\bibnamefont {Liang}}, \bibinfo {author} {\bibfnamefont {Q.}~\bibnamefont {Zou}}, \bibinfo {author} {\bibfnamefont {M.}~\bibnamefont {Fu}}, \bibinfo {author} {\bibfnamefont {A.~D.}\ \bibnamefont {Oyedele}}, \bibinfo {author} {\bibfnamefont {B.~G.}\ \bibnamefont {Sumpter}}, \bibinfo {author} {\bibfnamefont {Z.}~\bibnamefont {Liu}}, \bibinfo {author} {\bibfnamefont {Z.}~\bibnamefont {Gai}}, \bibinfo {author} {\bibfnamefont {K.}~\bibnamefont {Xiao}},\ and\ \bibinfo {author} {\bibfnamefont {A.-P.}\ \bibnamefont {Li}},\ }\bibfield  {title} {\bibinfo {title} {3d imaging and manipulation of subsurface selenium vacancies in {PdSe}$_{2}$},\ }\href@noop {} {\bibfield  {journal} {\bibinfo  {journal} {Physical Review Letters}\ }\textbf {\bibinfo {volume} {121}},\ \bibinfo {pages} {086101} (\bibinfo {year} {2018})}\BibitemShut {NoStop}%
\bibitem [{\citenamefont {Gr{\o}nvold}\ and\ \citenamefont {R{\o}st}(1957)}]{gronvold1957crystal}%
  \BibitemOpen
  \bibfield  {author} {\bibinfo {author} {\bibfnamefont {F.}~\bibnamefont {Gr{\o}nvold}}\ and\ \bibinfo {author} {\bibfnamefont {E.}~\bibnamefont {R{\o}st}},\ }\bibfield  {title} {\bibinfo {title} {The crystal structure of {PdSe}$_{2}$ and {PdS}$_{2}$},\ }\href@noop {} {\bibfield  {journal} {\bibinfo  {journal} {Acta Crystallographica}\ }\textbf {\bibinfo {volume} {10}},\ \bibinfo {pages} {329} (\bibinfo {year} {1957})}\BibitemShut {NoStop}%
\bibitem [{\citenamefont {Zhang}\ \emph {et~al.}(2024)\citenamefont {Zhang}, \citenamefont {Tian}, \citenamefont {Li}, \citenamefont {Yoon}, \citenamefont {Nelson}, \citenamefont {Li}, \citenamefont {Watanabe}, \citenamefont {Taniguchi}, \citenamefont {Smirnov}, \citenamefont {Kawakami} \emph {et~al.}}]{zhang2024quantum}%
  \BibitemOpen
  \bibfield  {author} {\bibinfo {author} {\bibfnamefont {Y.}~\bibnamefont {Zhang}}, \bibinfo {author} {\bibfnamefont {H.}~\bibnamefont {Tian}}, \bibinfo {author} {\bibfnamefont {H.}~\bibnamefont {Li}}, \bibinfo {author} {\bibfnamefont {C.}~\bibnamefont {Yoon}}, \bibinfo {author} {\bibfnamefont {R.~A.}\ \bibnamefont {Nelson}}, \bibinfo {author} {\bibfnamefont {Z.}~\bibnamefont {Li}}, \bibinfo {author} {\bibfnamefont {K.}~\bibnamefont {Watanabe}}, \bibinfo {author} {\bibfnamefont {T.}~\bibnamefont {Taniguchi}}, \bibinfo {author} {\bibfnamefont {D.}~\bibnamefont {Smirnov}}, \bibinfo {author} {\bibfnamefont {R.~K.}\ \bibnamefont {Kawakami}}, \emph {et~al.},\ }\bibfield  {title} {\bibinfo {title} {Quantum octets in high mobility pentagonal two-dimensional {PdSe}$_{2}$},\ }\href@noop {} {\bibfield  {journal} {\bibinfo  {journal} {Nature Communications}\ }\textbf {\bibinfo {volume} {15}},\ \bibinfo {pages} {761} (\bibinfo {year} {2024})}\BibitemShut {NoStop}%
\bibitem [{\citenamefont {Gao}\ \emph {et~al.}(2021)\citenamefont {Gao}, \citenamefont {Hu}, \citenamefont {Lu}, \citenamefont {Liu},\ and\ \citenamefont {Ni}}]{gao2021defect}%
  \BibitemOpen
  \bibfield  {author} {\bibinfo {author} {\bibfnamefont {L.}~\bibnamefont {Gao}}, \bibinfo {author} {\bibfnamefont {Z.}~\bibnamefont {Hu}}, \bibinfo {author} {\bibfnamefont {J.}~\bibnamefont {Lu}}, \bibinfo {author} {\bibfnamefont {H.}~\bibnamefont {Liu}},\ and\ \bibinfo {author} {\bibfnamefont {Z.}~\bibnamefont {Ni}},\ }\bibfield  {title} {\bibinfo {title} {Defect-related dynamics of photoexcited carriers in 2\text{D} transition metal dichalcogenides},\ }\href@noop {} {\bibfield  {journal} {\bibinfo  {journal} {Physical Chemistry Chemical Physics}\ }\textbf {\bibinfo {volume} {23}},\ \bibinfo {pages} {8222} (\bibinfo {year} {2021})}\BibitemShut {NoStop}%
\bibitem [{\citenamefont {{HQ Graphene}}(2025)}]{hqgraphene}%
  \BibitemOpen
  \bibfield  {author} {\bibinfo {author} {\bibnamefont {{HQ Graphene}}},\ }\href@noop {} {\bibinfo {title} {Hq graphene official website}},\ \bibinfo {howpublished} {\url{https://www.hqgraphene.com/}} (\bibinfo {year} {2025}),\ \bibinfo {note} {accessed: 2025-03-28}\BibitemShut {NoStop}%
\bibitem [{\citenamefont {Peli}\ \emph {et~al.}(2020)\citenamefont {Peli}, \citenamefont {Puntel}, \citenamefont {Kopic}, \citenamefont {Sockol}, \citenamefont {Parmigiani},\ and\ \citenamefont {Cilento}}]{peli2020time}%
  \BibitemOpen
  \bibfield  {author} {\bibinfo {author} {\bibfnamefont {S.}~\bibnamefont {Peli}}, \bibinfo {author} {\bibfnamefont {D.}~\bibnamefont {Puntel}}, \bibinfo {author} {\bibfnamefont {D.}~\bibnamefont {Kopic}}, \bibinfo {author} {\bibfnamefont {B.}~\bibnamefont {Sockol}}, \bibinfo {author} {\bibfnamefont {F.}~\bibnamefont {Parmigiani}},\ and\ \bibinfo {author} {\bibfnamefont {F.}~\bibnamefont {Cilento}},\ }\bibfield  {title} {\bibinfo {title} {Time-resolved vuv {ARPES} at 10.8 ev photon energy and {MHz} repetition rate},\ }\href@noop {} {\bibfield  {journal} {\bibinfo  {journal} {Journal of Electron Spectroscopy and Related Phenomena}\ }\textbf {\bibinfo {volume} {243}},\ \bibinfo {pages} {146978} (\bibinfo {year} {2020})}\BibitemShut {NoStop}%
\bibitem [{\citenamefont {Giannozzi}\ \emph {et~al.}(2009)\citenamefont {Giannozzi}, \citenamefont {Baroni}, \citenamefont {Bonini}, \citenamefont {Calandra}, \citenamefont {Car}, \citenamefont {Cavazzoni}, \citenamefont {Ceresoli}, \citenamefont {Chiarotti}, \citenamefont {Cococcioni}, \citenamefont {Dabo} \emph {et~al.}}]{giannozzi2009quantum}%
  \BibitemOpen
  \bibfield  {author} {\bibinfo {author} {\bibfnamefont {P.}~\bibnamefont {Giannozzi}}, \bibinfo {author} {\bibfnamefont {S.}~\bibnamefont {Baroni}}, \bibinfo {author} {\bibfnamefont {N.}~\bibnamefont {Bonini}}, \bibinfo {author} {\bibfnamefont {M.}~\bibnamefont {Calandra}}, \bibinfo {author} {\bibfnamefont {R.}~\bibnamefont {Car}}, \bibinfo {author} {\bibfnamefont {C.}~\bibnamefont {Cavazzoni}}, \bibinfo {author} {\bibfnamefont {D.}~\bibnamefont {Ceresoli}}, \bibinfo {author} {\bibfnamefont {G.~L.}\ \bibnamefont {Chiarotti}}, \bibinfo {author} {\bibfnamefont {M.}~\bibnamefont {Cococcioni}}, \bibinfo {author} {\bibfnamefont {I.}~\bibnamefont {Dabo}}, \emph {et~al.},\ }\bibfield  {title} {\bibinfo {title} {Quantum espresso: a modular and open-source software project for quantumsimulations of materials},\ }\href@noop {} {\bibfield  {journal} {\bibinfo  {journal} {Journal of physics: Condensed matter}\ }\textbf {\bibinfo {volume} {21}},\ \bibinfo {pages} {395502} (\bibinfo {year} {2009})}\BibitemShut {NoStop}%
\bibitem [{\citenamefont {Giannozzi}\ \emph {et~al.}(2017)\citenamefont {Giannozzi}, \citenamefont {Andreussi}, \citenamefont {Brumme}, \citenamefont {Bunau}, \citenamefont {Nardelli}, \citenamefont {Calandra}, \citenamefont {Car}, \citenamefont {Cavazzoni}, \citenamefont {Ceresoli}, \citenamefont {Cococcioni} \emph {et~al.}}]{giannozzi2017advanced}%
  \BibitemOpen
  \bibfield  {author} {\bibinfo {author} {\bibfnamefont {P.}~\bibnamefont {Giannozzi}}, \bibinfo {author} {\bibfnamefont {O.}~\bibnamefont {Andreussi}}, \bibinfo {author} {\bibfnamefont {T.}~\bibnamefont {Brumme}}, \bibinfo {author} {\bibfnamefont {O.}~\bibnamefont {Bunau}}, \bibinfo {author} {\bibfnamefont {M.~B.}\ \bibnamefont {Nardelli}}, \bibinfo {author} {\bibfnamefont {M.}~\bibnamefont {Calandra}}, \bibinfo {author} {\bibfnamefont {R.}~\bibnamefont {Car}}, \bibinfo {author} {\bibfnamefont {C.}~\bibnamefont {Cavazzoni}}, \bibinfo {author} {\bibfnamefont {D.}~\bibnamefont {Ceresoli}}, \bibinfo {author} {\bibfnamefont {M.}~\bibnamefont {Cococcioni}}, \emph {et~al.},\ }\bibfield  {title} {\bibinfo {title} {Advanced capabilities for materials modelling with {Quantum ESPRESSO}},\ }\href@noop {} {\bibfield  {journal} {\bibinfo  {journal} {Journal of physics: Condensed matter}\ }\textbf {\bibinfo {volume} {29}},\ \bibinfo {pages} {465901} (\bibinfo {year} {2017})}\BibitemShut {NoStop}%
\bibitem [{\citenamefont {Perdew}\ \emph {et~al.}(1996)\citenamefont {Perdew}, \citenamefont {Burke},\ and\ \citenamefont {Ernzerhof}}]{perdew1996generalized}%
  \BibitemOpen
  \bibfield  {author} {\bibinfo {author} {\bibfnamefont {J.~P.}\ \bibnamefont {Perdew}}, \bibinfo {author} {\bibfnamefont {K.}~\bibnamefont {Burke}},\ and\ \bibinfo {author} {\bibfnamefont {M.}~\bibnamefont {Ernzerhof}},\ }\bibfield  {title} {\bibinfo {title} {Generalized gradient approximation made simple},\ }\href@noop {} {\bibfield  {journal} {\bibinfo  {journal} {Physical Review Letters}\ }\textbf {\bibinfo {volume} {77}},\ \bibinfo {pages} {3865} (\bibinfo {year} {1996})}\BibitemShut {NoStop}%
\bibitem [{\citenamefont {Monkhorst}\ and\ \citenamefont {Pack}(1976)}]{monkhorst1976special}%
  \BibitemOpen
  \bibfield  {author} {\bibinfo {author} {\bibfnamefont {H.~J.}\ \bibnamefont {Monkhorst}}\ and\ \bibinfo {author} {\bibfnamefont {J.~D.}\ \bibnamefont {Pack}},\ }\bibfield  {title} {\bibinfo {title} {Special points for brillouin-zone integrations},\ }\href@noop {} {\bibfield  {journal} {\bibinfo  {journal} {Physical review B}\ }\textbf {\bibinfo {volume} {13}},\ \bibinfo {pages} {5188} (\bibinfo {year} {1976})}\BibitemShut {NoStop}%
\bibitem [{\citenamefont {Bart{\'o}k}\ and\ \citenamefont {Yates}(2019)}]{bartok2019regularized}%
  \BibitemOpen
  \bibfield  {author} {\bibinfo {author} {\bibfnamefont {A.~P.}\ \bibnamefont {Bart{\'o}k}}\ and\ \bibinfo {author} {\bibfnamefont {J.~R.}\ \bibnamefont {Yates}},\ }\bibfield  {title} {\bibinfo {title} {Regularized scan functional},\ }\href@noop {} {\bibfield  {journal} {\bibinfo  {journal} {The Journal of chemical physics}\ }\textbf {\bibinfo {volume} {150}} (\bibinfo {year} {2019})}\BibitemShut {NoStop}%
\bibitem [{\citenamefont {Sabatini}\ \emph {et~al.}(2013)\citenamefont {Sabatini}, \citenamefont {Gorni},\ and\ \citenamefont {De~Gironcoli}}]{sabatini2013nonlocal}%
  \BibitemOpen
  \bibfield  {author} {\bibinfo {author} {\bibfnamefont {R.}~\bibnamefont {Sabatini}}, \bibinfo {author} {\bibfnamefont {T.}~\bibnamefont {Gorni}},\ and\ \bibinfo {author} {\bibfnamefont {S.}~\bibnamefont {De~Gironcoli}},\ }\bibfield  {title} {\bibinfo {title} {Nonlocal van der waals density functional made simple and efficient},\ }\href@noop {} {\bibfield  {journal} {\bibinfo  {journal} {Physical Review B—Condensed Matter and Materials Physics}\ }\textbf {\bibinfo {volume} {87}},\ \bibinfo {pages} {041108} (\bibinfo {year} {2013})}\BibitemShut {NoStop}%
\bibitem [{\citenamefont {Heyd}\ \emph {et~al.}(2003)\citenamefont {Heyd}, \citenamefont {Scuseria},\ and\ \citenamefont {Ernzerhof}}]{heyd2003hybrid}%
  \BibitemOpen
  \bibfield  {author} {\bibinfo {author} {\bibfnamefont {J.}~\bibnamefont {Heyd}}, \bibinfo {author} {\bibfnamefont {G.~E.}\ \bibnamefont {Scuseria}},\ and\ \bibinfo {author} {\bibfnamefont {M.}~\bibnamefont {Ernzerhof}},\ }\bibfield  {title} {\bibinfo {title} {Hybrid functionals based on a screened coulomb potential},\ }\href@noop {} {\bibfield  {journal} {\bibinfo  {journal} {The Journal of chemical physics}\ }\textbf {\bibinfo {volume} {118}},\ \bibinfo {pages} {8207} (\bibinfo {year} {2003})}\BibitemShut {NoStop}%
\bibitem [{\citenamefont {Mostofi}\ \emph {et~al.}(2014)\citenamefont {Mostofi}, \citenamefont {Yates}, \citenamefont {Pizzi}, \citenamefont {Lee}, \citenamefont {Souza}, \citenamefont {Vanderbilt},\ and\ \citenamefont {Marzari}}]{mostofi2014updated}%
  \BibitemOpen
  \bibfield  {author} {\bibinfo {author} {\bibfnamefont {A.~A.}\ \bibnamefont {Mostofi}}, \bibinfo {author} {\bibfnamefont {J.~R.}\ \bibnamefont {Yates}}, \bibinfo {author} {\bibfnamefont {G.}~\bibnamefont {Pizzi}}, \bibinfo {author} {\bibfnamefont {Y.-S.}\ \bibnamefont {Lee}}, \bibinfo {author} {\bibfnamefont {I.}~\bibnamefont {Souza}}, \bibinfo {author} {\bibfnamefont {D.}~\bibnamefont {Vanderbilt}},\ and\ \bibinfo {author} {\bibfnamefont {N.}~\bibnamefont {Marzari}},\ }\bibfield  {title} {\bibinfo {title} {An updated version of wannier90: A tool for obtaining maximally-localised wannier functions},\ }\href@noop {} {\bibfield  {journal} {\bibinfo  {journal} {Computer Physics Communications}\ }\textbf {\bibinfo {volume} {185}},\ \bibinfo {pages} {2309} (\bibinfo {year} {2014})}\BibitemShut {NoStop}%
\end{thebibliography}%


\clearpage

\end{document}


\maketitle

\subsection{Crystal Structure}
PdSe$_2$ crystallizes in an orthorhombic structure with space group Pbca. Each Pd atom is coordinated by four Se atoms in a nearly square planar arrangement, while Se atoms form dimerized pairs, contributing to the material’s unique structural anisotropy. The unit cell consists of four Pd atoms and eight Se atoms, with interatomic distances of Pd–Se $\approx$ 2.44 \AA\ and Se–Se $\approx$ 2.36 \AA. The layered structure gives rise to strong $k_z$ dispersion (more details in the following sections), influencing its electronic properties across the Brillouin zone.
\begin{figure}{}
\centering
\includegraphics[width=0.7\textwidth]{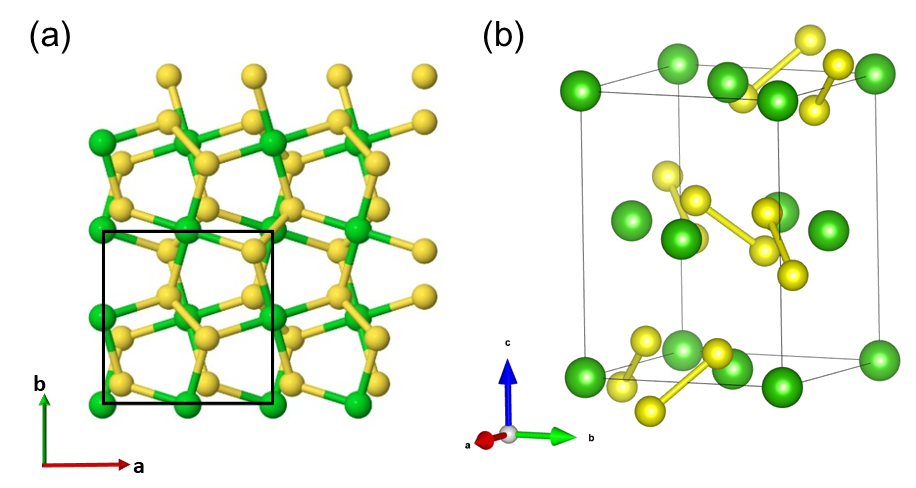}
\caption{(a) Top view of a PdSe${_2}$ bilayer. (b) Crystal structure of the PdSe${_2}$ unit cell. Green and yellow spheres represent Pd and Se atoms, respectively.}
\label{SF1}
\end{figure}

\subsection{The Expected k-resolved Density of States (k-DOS) for S- and P-polarized Light} 
\begin{figure}{}
\centering
\includegraphics[width=1\textwidth]{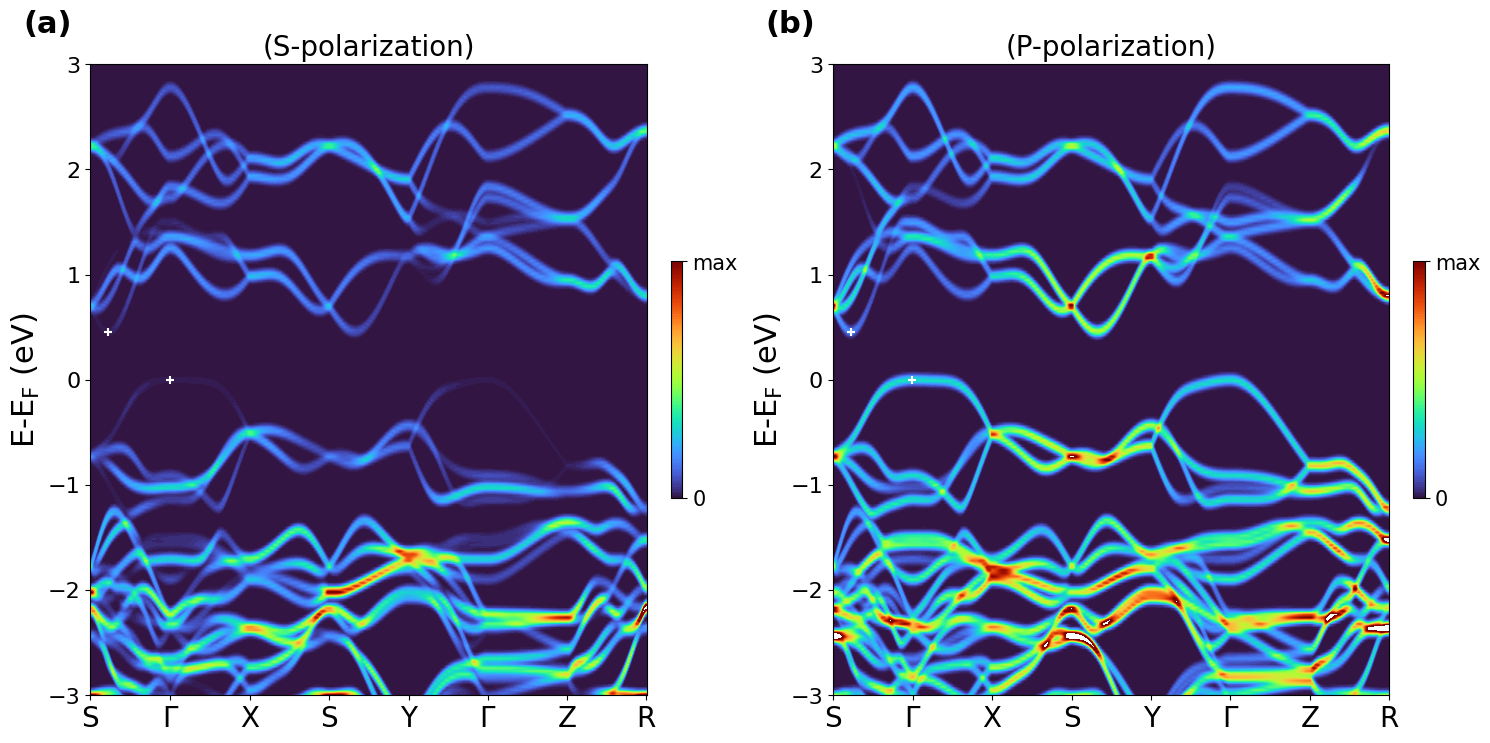}
\caption{The calculated density of states (DOS) of bulk PdSe$_2$ is presented for both S- and P-polarized light. For S-polarization (a), the Se $p_y$ and Pd $d_{xy}$, $d_{yz}$ orbitals are considered, while for P-polarization (b), the Se $p_x$, $p_z$ and Pd $d_{z^2}$, $d_{x^2-y^2}$, and $d_{zx}$ orbitals are taken into account. The white plus marks indicate the positions of the upper valence band and lower conduction band.}
\label{SF3}
\end{figure}
The use of S- and P-polarized light provides a selective method for probing electronic states based on their symmetry and orbital character, offering a deeper understanding of band dispersion and orbital interactions.\\
In the case of S-polarized light (perpendicular to the plane of incidence), the electric field is oriented perpendicular to the surface, primarily exciting in-plane orbitals. For PdSe$_2$, this includes the Se $p_y$ and Pd $d_{xy}$, $d_{yz}$ orbitals, as reported in reference \cite{gu2022low}. The total contribution of these orbitals is illustrated in Fig.~\ref{SF3}. As a result, the calculated k-DOS predominantly reflects electronic states with strong in-plane components, providing insights into bonding and non-bonding states in 2D materials. S-polarized light is particularly effective for distinguishing between electronic bands that may otherwise be difficult to separate in unpolarized measurements.\\
In contrast, P-polarized light (parallel to the plane of incidence) has both in-plane and out-of-plane electric field components, enabling the excitation of states with out-of-plane contributions. In the case of PdSe$_2$, this includes the Se $p_x$, $p_z$ and Pd $d_{z^2}$, $d_{zx}$ and $d_{x^2-y^2}$ orbitals, as shown in Fig.~\ref{SF3}. In this configuration, the calculated k-DOS emphasizes electronic states with vertical components, shedding light on interlayer interactions, surface states, and hybridized bands.\\
As shown from the calculated band structure for both S- and P-polarizations, it is evident that the spectral weight of the top valence band (VB) strongly depends on the light polarization. Specifically, the top VB is highly sensitive to P-polarization when the energy range $(E - E{\mathrm{_F}})$ spans from -0.5 to 0 eV. Consequently, the spectral weight under P-polarized light is significantly larger than that under S-polarized light, which accounts for the giant linear dichroism observed in previous experimental results. These findings provide a comprehensive understanding of the orbital character of both the valence and conduction bands, highlighting the substantial orbital-induced anisotropy in this material. This positions PdSe$_2$ as a novel, intrinsically orbital-engineered material with exceptional anisotropy, making it ideal for polarization-dependent optical applications.

\subsection{Background Subtraction and Fitting Procedure}
\begin{figure}[h!]
    \centering
    \includegraphics[width=0.7\textwidth]{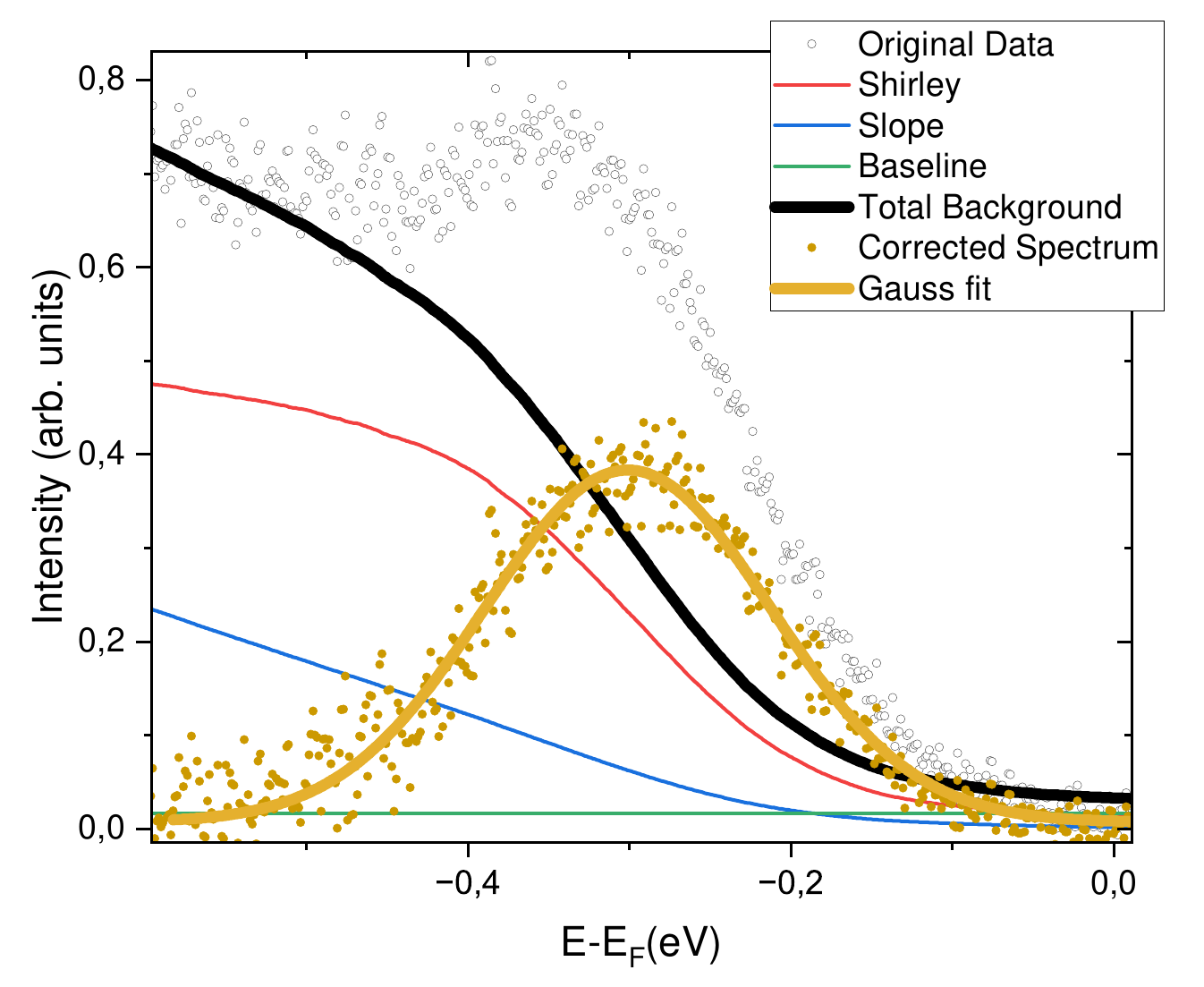}
    \caption{Background subtraction procedure applied to the valence band of PdSe$_2$. The total background (black curve) is decomposed into Shirley-Sherwood (red), slope (blue), and baseline (green) components. The black dots represent the raw experimental data and the brown dots are subtracted data and their Gaussian fit to derive the VB EDC.}
    \label{SF4}
\end{figure}

In order to accurately analyze the valence band structure of PdSe$_2$ near the Fermi level, it is crucial to remove inelastic background contributions from the measured photoemission spectrum. The total background is modeled as the sum of three components: (i) a Shirley-Sherwood background, which accounts for inelastic electron scattering events, (ii) a slope background, which represents the loss tail due to electron energy dissipation, and (iii) a constant baseline offset. This decomposition follows the active-background method used in previous studies to optimize the spectral fitting~\cite{shirley1972high,tougaard1997universality}.\\
The total background $B(E)$ is defined as:
\begin{equation}
B(E) = B_{\text{Shirley}}(E) + B_{\text{slope}}(E) + B_{\text{baseline}}.
\end{equation}\\
The contribution of each component is determined through iterative fitting. The iterative process ensures that $B_{\text{Shirley}}(E)$ correctly accounts for inelastic scattering contributions, particularly in regions where spectral weight accumulates due to electron interactions.
The slope background, which accounts for the electron energy loss tail, is modeled following an empirical formulation:
\begin{equation}
\frac{dB_{\text{slope}}(E)}{dE} = -k_{\text{slope}} \int_{E}^{E+\Delta} \left[ I(E') - I_{\text{right}} \right] dE',
\end{equation}
where $k_{\text{slope}}$ is a parameter that quantifies the strength of the background slope, and $\Delta$ defines the range over which the intensity is integrated. This functional form is motivated by inelastic electron transport theory, which describes how electrons lose energy via scattering events as they escape the material. The value of $k_{\text{slope}}$ is found through optimization and, in this study, closely matches the initial slope of the function, as predicted by Tougaard-Sigmund’s electron transport theory~\cite{tougaard1982influence}.\\
The final term in the background is a small constant offset:
\begin{equation}
B_{\text{baseline}} = c_0,
\end{equation}
where $c_0$ is extracted from the high-energy tail of the spectrum.\\
Physically, the Shirley-Sherwood component captures inelastic scattering events that redistribute spectral weight, the slope background models additional energy losses due to electron transport effects, and the baseline accounts for any instrumental or experimental offsets. This fitting method is justified as it correctly models the non-trivial energy-dependent background contributions observed in high-resolution ARPES spectra.\\
Fig.~\ref{SF4} shows the decomposition of the total background (magenta) into its constituent components. The subtraction of this background from the raw spectrum isolates the intrinsic spectral features of PdSe$_2$, enabling accurate determination of the valence band maximum.
\subsection*{Fitting Procedure of the Valence Band}

The VB dynamics is analyzed using a phenomenological model consisting of three distinct contributions attributed to band gap renormalization (BGR), valence band depletion (Dep), and surface photovoltage (SPV). Each of these processes is assumed to affect the VB lineshape while occurring with a characteristic timescale and contribute differently across energy regions because of the Gaussian profile of the VB. To describe the time evolution of the photoemission signal, we employ a model that combines rise and decay dynamics for BGR and Dep, and a single exponential decay for the SPV component. As clarified in the main text, SPV's rise time is affected by the TR-ARPES experiment and is prolonged, occurring at approximately 10 ps at negative delays (refer to Fig.~4d in the main manuscript). Hence, only its decay time is incorporated into the model. The time-dependent photoemission intensity in the $n^\text{th}$ energy region is given by:
\begin{equation}
\begin{split}
PI_{(n)}(t) = \bigg\{\bigg[ & A_{(n)}^\mathrm{BGR} \left( 1 - e^{-t/\tau_\mathrm{r}^\mathrm{BGR}} \right) e^{-t/\tau_\mathrm{d}^\mathrm{BGR}} \\
& + A_{(n)}^\mathrm{Dep} \left( 1 - e^{-t/\tau_\mathrm{r}^\mathrm{Dep}} \right) e^{-t/\tau_\mathrm{d}^\mathrm{Dep}} \\
& + A_{(n)}^\mathrm{SPV} e^{-t/\tau_\mathrm{d}^\mathrm{SPV}} \bigg] \times \mathcal{H}(t) \bigg\} \otimes \mathcal{R}(t),
\end{split}
\label{eq:fit_dynamics}
\end{equation}

Here, $A_{(n)}^i$ denotes the amplitude of the $i^\text{th}$ process (BGR, Dep, SPV) in the $n^\text{th}$ energy region. The parameters $\tau_\mathrm{r}^i$ and $\tau_\mathrm{d}^i$ represent the rise and decay time constants for the respective processes. All energy regions share the same global set of time constants $\tau^i$, while the amplitudes $A_{(n)}^i$ vary with energy to account for region-specific contributions.\\
The function $\mathcal{R}(t)$ is a Gaussian instrumental response, which accounts for the finite time resolution of the experimental setup and broadens the observed signal accordingly. The Heaviside function $\mathcal{H}(t)$ defines the onset of photoexcitation, implemented through the error function representation. Using this minimal and physically motivated set of assumptions, we successfully reproduce the energy-dependent spectral weight dynamics of the VB and extract time constants associated with the key ultrafast processes. The extracted fitting parameters are visiulized in Fig.~3c of the main manuscript and detailed in the table below.
\\
\begin{table}[h!]
\centering
\begin{tabular}{|c|c|c|c|}
\hline
\textbf{Parameter} & \textbf{Region 3} & \textbf{Region 2} & \textbf{Region 1} \\
\hline
Broadening Amp & $0 \pm 0.8$ & $1.59 \pm 0.60$ & $4.64 \pm 0.97$ \\
$t_{r}^{Br}$ ($<$ response time) & $<0.35$ & $<0.35$ & $<0.35$ \\
$t_{d}^{Br}$ & $0.68 \pm 0.35$ & $0.68 \pm 0.35$ & $0.68 \pm 0.35$ \\
Depletion Amp & $-3.62 \pm 1.35$ & $-1.61 \pm 0.73$ & $-1.49 \pm 1.02$ \\
$t_{d}^{Dep}$ & $2.76 \pm 0.77$ & $2.76 \pm 0.77$ & $2.76 \pm 0.77$ \\
$t_{r}^{Dep}$ & $0.62 \pm 0.41$ & $0.62 \pm 0.41$ & $0.62 \pm 0.41$ \\
SPV Amp & $5.54 \pm 0.83$ & $9.27 \pm 1.32$ & $4 \pm 0.63$ \\
$t_{d}^{SPV}$ & $59.75 \pm 12.73$ & $59.75 \pm 12.73$ & $59.75 \pm 12.73$ \\
\hline
\end{tabular}
\caption{Obtained parameters for three different regions of VB (as shown in Fig. 3a in the main manuscript) in PdSe$_2$. The parameters include the amplitude of broadening or band gap renormalization (BGR) (Broadening Amp), the amplitude of depletion (Depletion Amp), the rise time for broadening (or BGR) ($t_{r}^{Br}$), the decay time for broadening (or BGR) ($t_{d}^{Br}$), the decay time for depletion ($t_{d}^{Dep}$), the rise time for depletion ($t_{r}^{Dep}$), the amplitude of surface photovoltage (SPV Amp), and the decay time for SPV ($t_{d}^{SPV}$).} 
\label{tab:params}
\end{table}

\subsection*{Determination of the Indirect Band Gap of PdSe$_2$}
\begin{figure}{}
\centering
\includegraphics[width=1\textwidth]{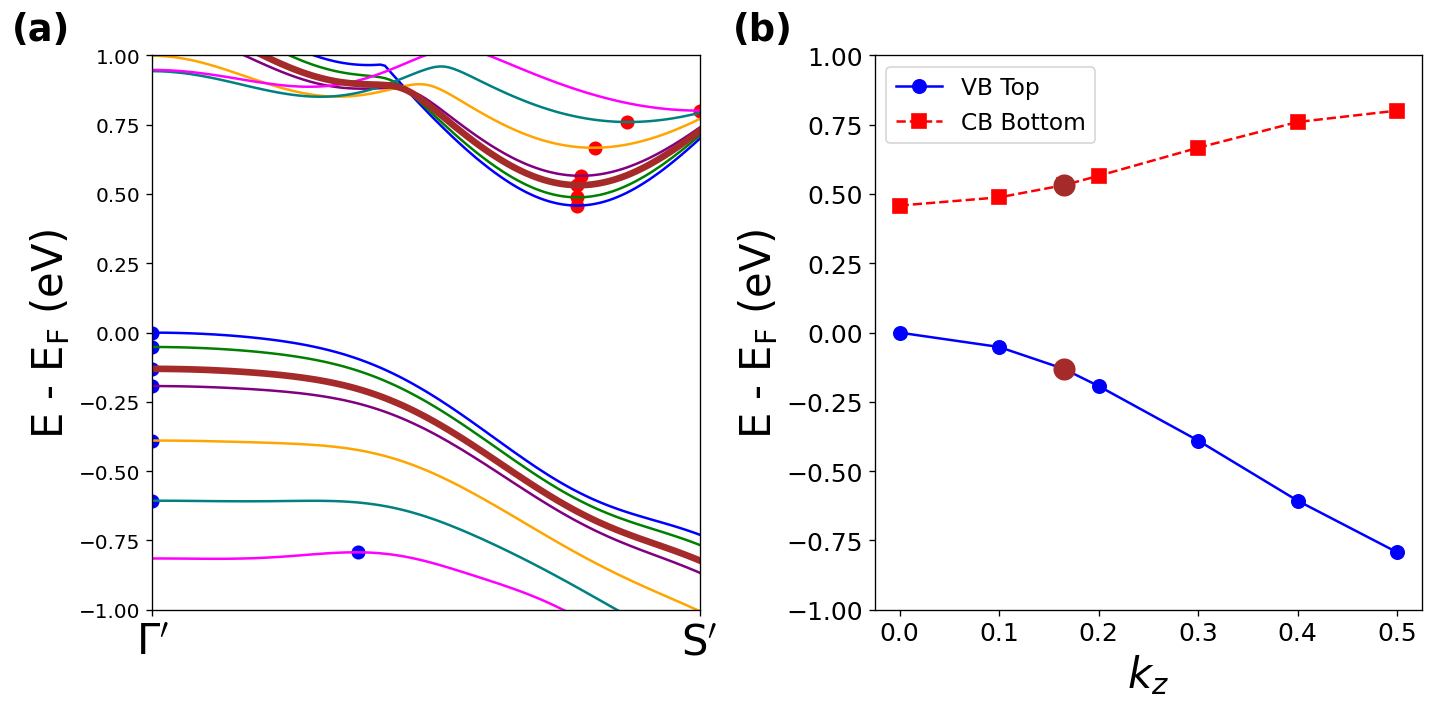}
\caption{(a) Close-up of density functional theory (DFT) calculations for the highest valence band (VB) and the lowest conduction band (CB) along the $\Gamma^\prime$-S$^\prime$ axis at seven distinct $K_z$ values. The bands are plotted for the following $K_z$ values: $K_z = 0$ (black), $K_z = 0.1$ (blue), $K_z = 0.165$ (brown), $K_z = 0.4$ (green), $K_z = 0.6$ (red), $K_z = 0.8$ (light blue), and $K_z = 1$ (purple). The blue and red bullet marks indicate the positions of the VBM and CBM, respectively. (b) Calculated $K_z$ dispersion of the bottom CB and the top VB. The brown dots correspond to the estimated VB and CB dispersion for $K_z = 0.165$.}
\label{SF5}
\end{figure}
To determine the indirect band gap of PdSe$_2$, we combine TR-ARPES measurements with the relative $k_z$-dependent dispersion trends from DFT. While DFT typically underestimates absolute band gaps, it captures the variation of the gap along the $k_z$ direction reliably, as depicted in Fig.~\ref{SF5}. We note that the trends of the band edge positions at the $\Gamma^\prime$ and S$^\prime$ points with changing $k_z$ were found to be similar across all levels of theory (PBE, SCAN+rVV10, and HSE, see Section Methods, main text).\\
The probe photon energy used in the TR-ARPES experiment is 10.80 eV. Given the known periodicity of $\Gamma \rightarrow Z \rightarrow \Gamma$ transitions in photon-energy-dependent ARPES (approximately 41.5 eV per full period), this photon energy corresponds to a normalized out-of-plane momentum of $k_z = 0.162$, where $k_z = 0$ represents the $\Gamma$ point and $k_z = 0.5$ the $Z$ point.\\
We define:
\[
\text{BG}_{\text{TR-ARPES}} = \text{BG}^{\text{DFT}}_{0.162} + dE_g,
\]
where $\text{BG}_{\text{TR-ARPES}}$ is the experimentally measured band gap at $k_z = 0.162$, $\text{BG}^{\text{DFT}}_{0.162}$ is the DFT band gap at the same $k_z$, and $dE_g$ is the unknown DFT band gap offset.\\
We compute the $k_z$-dependent ratio from DFT:
\[
r = \frac{\text{BG}^{\text{DFT}}_{0.162}}{\text{BG}^{\text{DFT}}_{0}}.
\]
This ratio is independent of $dE_g$, assuming the same global error applies across all $k_z$.\\
Solving for $dE_g$:
\[
dE_g = \text{BG}_{\text{TR-ARPES}} - \text{BG}^{\text{DFT}}_{0.162},
\]
we then estimate the corrected indirect band gap at the $\Gamma$ point as:
\[
\text{BG}_{\Gamma} = \text{BG}^{\text{DFT}}_0 + dE_g = \text{BG}_{\text{TR-ARPES}} + (1 - r) \cdot \text{BG}^{\text{DFT}}_0.
\]\\
Alternatively, rearranging the expressions, the final corrected gap can be expressed as:
\[
\text{BG}_{\Gamma} = \text{BG}_{\text{TR-ARPES}} \cdot \frac{1}{r},
\]
if DFT predicts a vanishing gap at $\Gamma$ (i.e., $\text{BG}^{\text{DFT}}_0 = 0$), which is the case in some calculations.

\vspace{0.3em}

From TR-ARPES measurements, the valence band maximum is located at $-0.26$ eV and the conduction band minimum at $+0.54$ eV, yielding:
\[
\text{BG}_{\text{TR-ARPES}} = 0.54 - (-0.26) = 0.80 \, \text{eV}.
\]
From DFT at $k_z = 0.162$, we have:
\[
\text{BG}^{\text{DFT}}_{0.162} = 0.5321 - (-0.1307) = 0.6628 \, \text{eV},
\]
and at $\Gamma$:
\[
\text{BG}^{\text{DFT}}_0 = 0.4585 \, \text{eV}.
\]
The DFT ratio is then:
\[
r = \frac{0.6628}{0.4585} \approx 1.446.
\]
Using the inverse scaling:
\[
\text{BG}_{\Gamma} = \frac{0.80}{1.446} \approx 0.55 \, \text{eV}.
\]
This procedure enables a robust extraction of the true indirect band gap of PdSe$_2$, correcting the DFT results using experimentally anchored data while relying only on relative $k_z$-dependent DFT trends.

\begin{figure}[h!]
    \centering
    \includegraphics[width=0.4\textwidth]{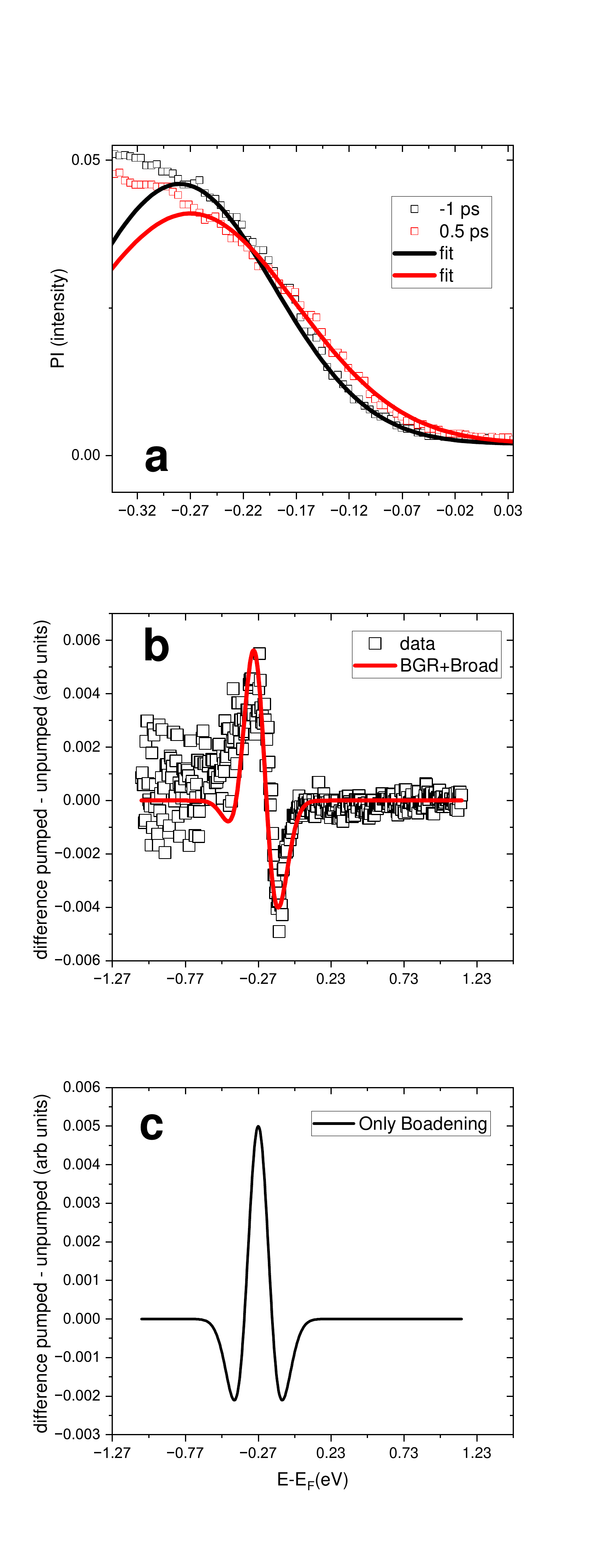}
    \caption{(a) EDC of VB before (black dots) and after (red dots) photoexcitation, with Gaussian fits depicted by solid lines indicating the best fit of the difference between Gaussian functions to the pumped-unpumped data. (b) Difference between pumped and unpumped EDCs, with a fitting curve illustrating both broadening and VB shift or BGR. (c) Hypothetical difference of two EDCs if only VB broadening occurred, demonstrating the actual presence of BGR as opposed to the observed data.}
    \label{SF6}
\end{figure}
\subsection*{Analysis of EDCs Before and After Photoexcitation} 
The assessment of VB dynamics before and after photoexcitation presents significant challenges due to the substantial background signal. Despite using various background subtraction methods described in this manuscript, uncertainties persist, primarily because the VB energy shifts and broadening are within tens of meV. Such small changes render straightforward background subtraction suboptimal. We assume that the background primarily arises from probe photoemission and that the pump's impact on this background is negligible\cite{hedayat2018surface}. This assumption is verified by observing that the low-energy tails of both pumped and unpumped data overlap perfectly, as shown in Fig.~\ref{SF6}a. \\
To isolate the effects of photoexcitation on the VB, we subtract the EDC of the unpumped state from the pumped state, thereby eliminating the influence of the background. Fig.~\ref{SF6}b illustrates the resulting difference between the two EDCs. We model both the pumped and unpumped VB using Gaussian profiles characterized by amplitude (A), center energy (\(x_c\)), and full width at half maximum (FWHM, \(w\)). The difference between these Gaussian profiles is fitted to the differential data shown in Fig.~\ref{SF6}b.\\
Subsequently, we fit the higher energy edges of both pumped and unpumped data simultaneously, depicted by solid lines in Fig.~\ref{SF6}a. The values obtained for broadening and VB shift range from 10 to 30 meV, comparable with the energy resolution of the time-resolved ARPES (TR-ARPES) apparatus. Fig.~\ref{SF5}c contrasts the expected difference if only VB broadening occurred, without band gap renormalization, against the observed difference which suggests both phenomena are present. This comparison further supports the occurrence of band gap renormalization, as the shape of the pumped-unpumped difference does not align with the scenario of mere VB broadening.

\subsection{Raman Measurements on PdSe\textsubscript{2}}
\begin{figure}{}
\centering
\includegraphics[width=1\textwidth]{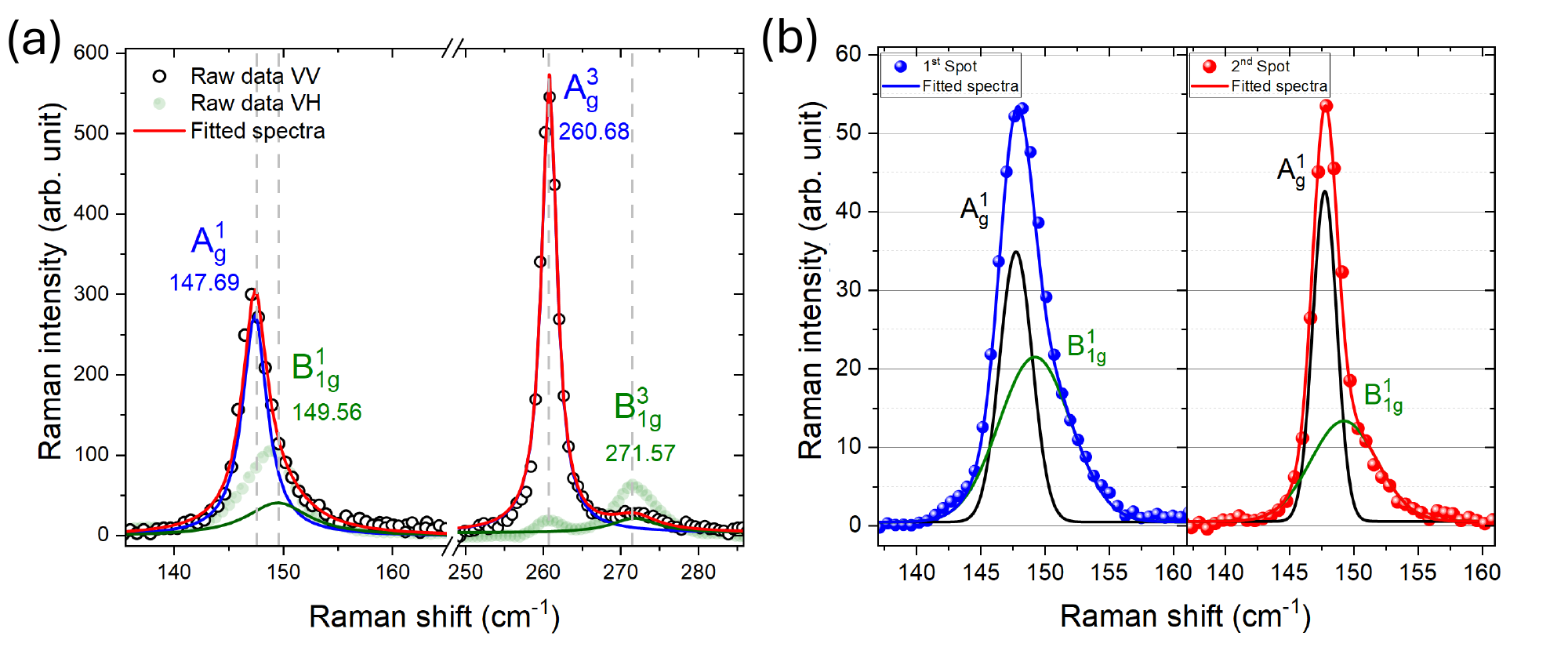}
\caption{(a) Raman spectra of bulk PdSe\textsubscript{2} in both parallel (VV) and cross-polarized (VH) configurations, with experimental data shown as symbols and corresponding fits as solid lines. The prominent Raman-active phonon modes, $A^{1}_{g}$ and $A^{3}_{g}$, along with their characteristic asymmetric line shapes, originating from the presence of the forbidden $B^{1}_{1g}$ mode scattering in VV scattering geometry, are highlighted. The right panel (b) compares the $A^{1}_{g}$ to $B^{1}_{1g}$ intensity ratio at two distinct microscale locations on the sample.}
\label{SFR}
\end{figure}

Our Raman measurements on PdSe$_2$ revealed the presence of $B_{1g}$ modes in the parallel polarization geometry, where such modes are forbidden by symmetry selection rules (see Fig.~\ref{SFR}(a)). Spatial mapping across micron-scale regions of the sample showed slight variations in the $A^1_{g}/B^1_{1g}$ intensity ratio (see Fig.~\ref{SFR}(b)). The observed violation of symmetry selection rules is therefore attributed to defect--phonon coupling, providing independent evidence of a significant intrinsic defect landscape in our samples. Recent experimental and theoretical studies have demonstrated that intercalation can modify this ratio in PdSe$_2$, lending further support to our observations~\cite{jena2023evidence}.\\
The noticeable variation in the $A^1_{g}/B^1_{1g}$ intensity ratio across different locations suggests a slight local inhomogeneity, likely originating from spatial variations in defect density. This interpretation is further supported by the corresponding increase in the full width at half maximum (FWHM), as higher defect concentrations typically enhance phonon scattering and broaden Raman peaks. The probe spot size in our TR-ARPES measurements ($\sim$200\,$\mu$m) is approximately two orders of magnitude larger than that used in the Raman experiments. As a result, the ARPES data represent an average over the spatial inhomogeneity observed in the micro-Raman spectra.

\subsection{Defect States in DFT Calculations}
\begin{figure}{}
\centering
\includegraphics[width=1\textwidth]{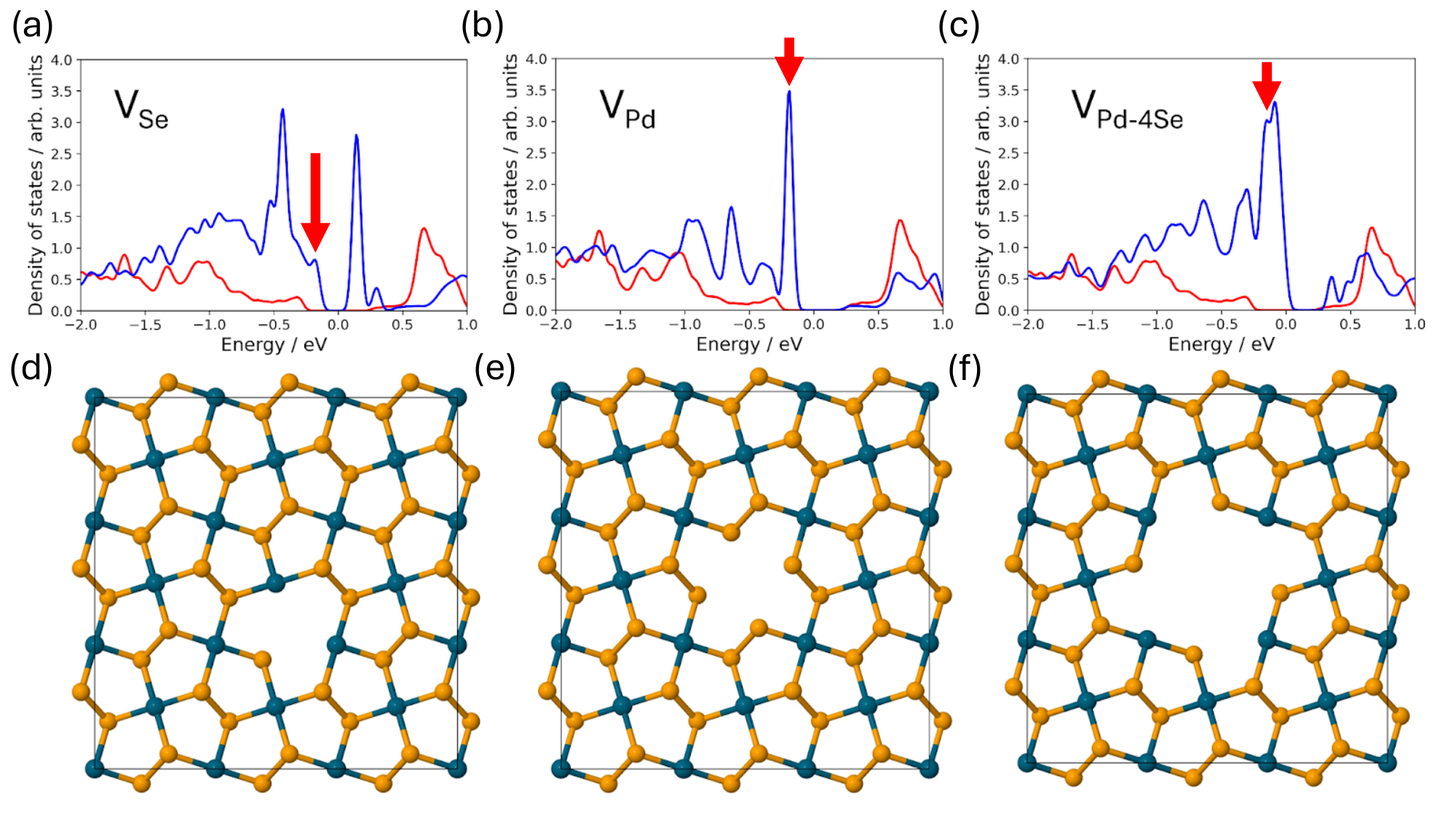}
\caption{Calculated density of states (DOS) (a--c) of PdSe\textsubscript{2} with (blue) and without (red) defects, shown for different vacancy types: Se vacancy (V\textsubscript{Se}) in (a), Pd vacancy (V\textsubscript{Pd}) in (b), and combined Pd--Se vacancy (V\textsubscript{Pd--4Se}) in (c), and their corresponding atomic structures (d--f). Dark turquoise and orange spheres represent Pd and Se atoms, respectively.}
\label{SF8}
\end{figure}

We examined a defect-free PdSe\textsubscript{2} structure and three probable point defects identified in previous studies: a selenium vacancy (V\textsubscript{Se}), a palladium vacancy (V\textsubscript{Pd}), and a complex vacancy consisting of one Pd atom and its four neighboring Se atoms (V\textsubscript{Pd--4Se}), as depicted in Fig.~\ref{SF8}. Each defect was introduced in the surface layer of a 3\,$\times$\,3 in-plane supercell with three layers along the \emph{c}-axis (162 atoms in total). Neighboring atoms were relaxed until forces were below $10^{-3}$\,eV\,\AA$^{-1}$. A vacuum spacing of 20\,\AA{} was included along the \emph{c}-axis to ensure electronic decoupling between slabs. Because monolayer and bilayer results (as presented in refs.~\cite{jena2023evidence,nguyen20183d}) differed significantly from thicker slabs (as in ref.~\cite{fu2019defects}), we selected a minimum thickness of three layers. Wavefunction projections onto atomic orbitals were used to identify defect-related peaks in the density of states below the Fermi level. As expected, the dominant contributions originated from atoms with broken bonds to the missing species, primarily Se \emph{p}-orbitals with minor Pd \emph{d}-orbital contributions. Projections on atoms near the supercell edges and in the lower layers were largely unaffected by the defects, supporting the adequacy of the chosen supercell size.\\
Our observations of the states below $E_\mathrm{F}$, in the ARPES measurments, are in excellent agreement with these advanced DFT calculations~\cite{ciocys2019tracking,jena2023evidence,nguyen20183d}, indicating that defect vacancies preferentially accumulate at the surface and generate mid-gap states consistent with our findings. This behavior has also been confirmed experimentally~\cite{fu2019defects}.

\section{Estimation of Surface Defect Density from SPV}
The estimation of surface defect density from the observed SPV is based on Poisson’s equation under the depletion approximation. In this framework, the space charge region near the semiconductor surface is assumed to be depleted of mobile carriers, so the charge density is set by ionized dopants. Solving Poisson’s equation yields the relationship:
\begin{equation}
    Q_s = \sqrt{2\varepsilon q N_D \Delta V}
\end{equation}
where $Q_s$ is the surface charge density, $\varepsilon$ is the dielectric constant of PdSe$_2$ reasonably assumed $\varepsilon_r \approx 15$, $q$ is the elementary charge, $N_D$ is the bulk doping density (assumed $10^{19}$--$10^{20}$~cm$^{-3}$~\cite{abbas2025first}), and $\Delta V$ is the measured SPV (67~meV). The corresponding surface defect density is then given by:
\begin{equation}
    N_{sd} = \frac{Q_s}{q}
\end{equation}
For the parameters relevant to PdSe$_2$, this yields $N_{sd} \sim 10^{12}$--$10^{13}$~cm$^{-2}$, consistent with the values required for robust SPV effects~\cite{kronik1999surface}. Further experimental studies and refined material parameters will help to improve the accuracy of this estimate.

\subsection{Fermi Level Calibration Using Gold as a Reference in ARPES Measurements}

\begin{figure}{}
\centering
\includegraphics[width=0.56\textwidth]{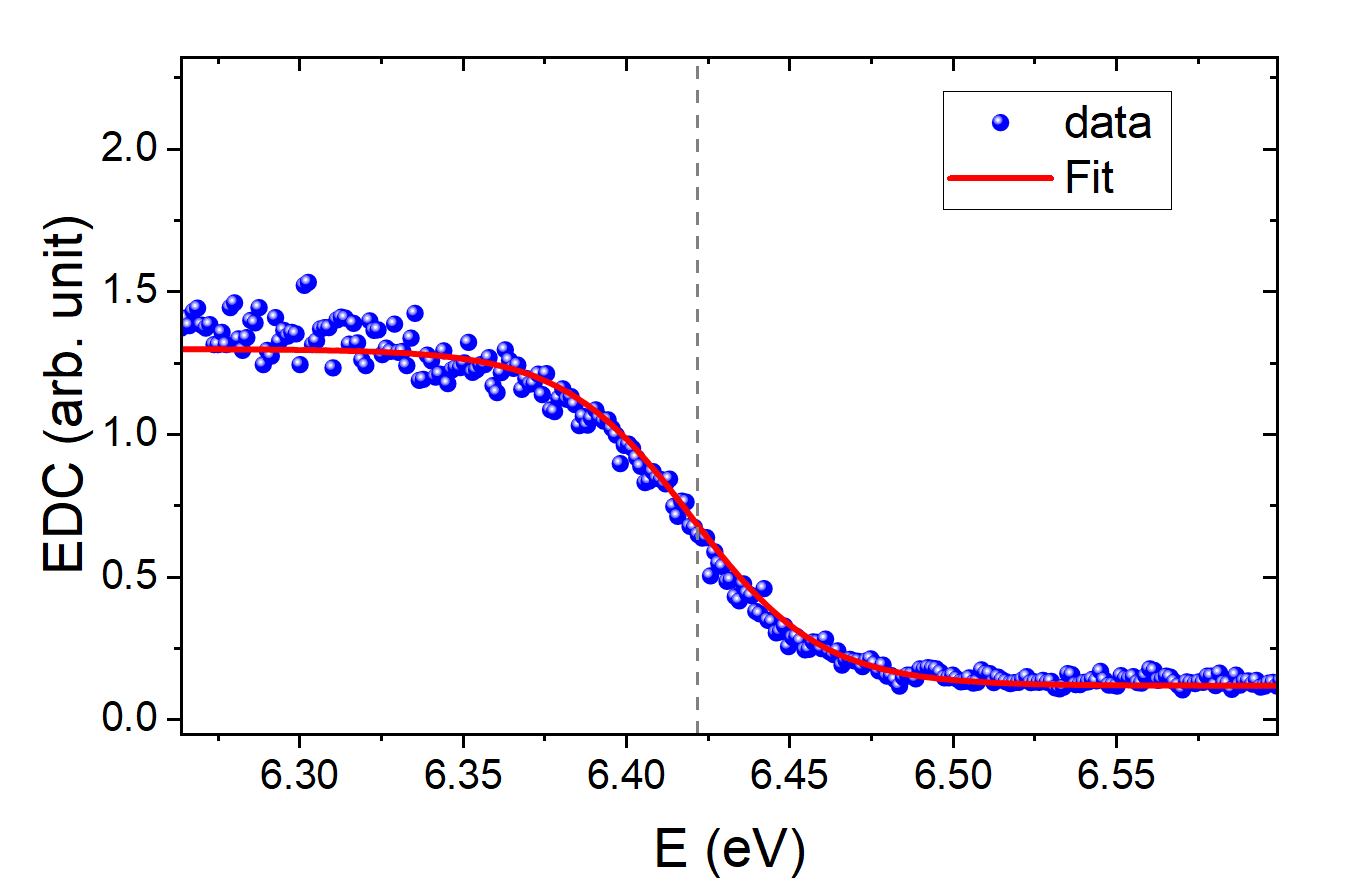}
\caption{EDC of the ARPES data cut for gold. The red line represents the Fermi-Dirac distribution fitting, with the Fermi energy determined to be about 6.42 eV. }
\label{SF9}
\end{figure}

In our ARPES measurements, polycrystalline gold (Au) was used as a reference to determine the Fermi level of the sample. Gold's well-defined Fermi edge provides a precise calibration for the Fermi level of the ARPES maps. By fitting the Fermi edge in gold's ARPES spectra, we determined the Fermi level of the sample to be about 6.42 eV as shown in Fig.~\ref{SF9}. This calibration ensures the accuracy and consistency of all subsequent ARPES measurements.
The Fermi-Dirac distribution used to determine the Fermi level is:

\[
f(E) = \frac{1}{e^{(E - E_F) / k_B T} + 1}
\]
where \( E_F \) is the Fermi level, \( k_B \) is Boltzmann's constant, and \( T \) is the absolute temperature.